\documentclass{aa}
\usepackage{graphicx}
\usepackage{txfonts}
\usepackage{xcolor}
\usepackage{mathrsfs}
\usepackage{amssymb,amsmath}
\usepackage{amsmath}	
\usepackage{amssymb}	
\DeclareMathAlphabet{\mathbi}{OT1}{ptm}{bx}{it}
\SetMathAlphabet\mathbi{bold}{OT1}{ptm}{bx}{it}

\DeclareMathOperator{\EX}{\mathbb{E}}
\usepackage{lipsum}
\usepackage{stfloats}

\newcommand{\bd}[1]{\mbox{\boldmath $#1$}}
\usepackage{hyperref}
\hypersetup{
    colorlinks=true,
    citecolor=blue,
    linkcolor=blue,
    filecolor=magenta,      
    urlcolor=cyan,
}

\usepackage{subfig}

\newcommand{\RR}[1]{\mathbb{R}^{#1}} 

\DeclareMathOperator*{\argmax}{arg\,max}

\newcommand{\appropto}{\mathrel{\vcenter{
  \offinterlineskip\halign{\hfil$##$\cr
    \propto\cr\noalign{\kern2pt}\sim\cr\noalign{\kern-2pt}}}}}
    
\begin{document} 

   \title{A Gaussian process cross-correlation approach to time delay estimation in active galactic nuclei.\thanks{The code can be downloaded from: \url{https://github.com/HITS-AIN/GPCC.jl/}. Also indexed under \url{https://ascl.net/2303.006}.} \fnmsep \thanks{Instructions and specific examples used in this paper can be found in: \url{https://github.com/HITS-AIN/GPCCpaper}}}

   \titlerunning{Gaussian Process Cross Correlation}


   \author{F. Pozo Nu\~nez\href{https://orcid.org/0000-0002-6716-4179}{\includegraphics[scale=0.5]{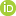}}
          \and
          N. Gianniotis
         \and
          K. L. Polsterer
        }
    
   \institute{Astroinformatics, Heidelberg Institute for Theoretical Studies, Schloss-Wolfsbrunnenweg 35, 69118 Heidelberg, Germany \\
              \email{francisco.pozon@gmail.com}}

   \date{Received ??, 2022; accepted ??, 2022}

 \abstract{We present a probabilistic cross-correlation approach to estimate time delays in the context of reverberation mapping (RM) of Active Galactic Nuclei (AGN).}
 {We reformulate the traditional interpolated cross-correlation method as a statistically principled model that delivers a posterior distribution for the delay.}
 {The method employs Gaussian processes as a model for observed AGN light curves. We describe the mathematical formalism and demonstrate the new approach using both simulated light curves and available RM observations.}
 {The proposed method delivers a posterior distribution for the delay that accounts for observational noise and the non-uniform sampling of the light curves. This feature allow us to fully quantify its uncertainty and propagate it to subsequent calculations of dependent physical quantities, e.g., black hole masses. It delivers out-of-sample predictions, which enables us to subject it to model selection and it can calculate the joint posterior delay for more than two light curves. }
 {Because of the numerous advantages of our reformulation and the simplicity of its application, we anticipate that our method will find favour not only in the specialised community of RM, but in all fields where cross-correlation analysis is performed. We provide the algorithms and examples of their application as part of our \textit{Julia} GPCC package.}


   \keywords{galaxies: active --galaxies: quasars
          --galaxies: nuclei --galaxies: Seyfert
          --methods: statistical --methods: observational
               }

   \maketitle
   
%
%

\section{Introduction}

Reverberation mapping (RM; \citealt{1973ApL....13..165C}; \citealt{1982ApJ...255..419B}; \citealt{1986ApJ...305..175G}; \citealt{2004ApJ...613..682P}) relies on observed variability to measure the time delay ($\tau$) between changes in the continuum of the accretion disk (AD) and different regions in active galactic nuclei (AGN).
In particular, the delay provides an estimate of the distance between AD continuum regions (or the AD size; $R_{\rm AD } \propto c\cdot\tau_{\rm AD }$, $c$ is the speed of light, e.g.,  \citealt{2019MNRAS.490.3936P}; \citealt{2020ApJ...896....1C}), the broad emission line region (BLR) clouds to AD (or the BLR size; $R_{\rm BLR} \propto c\cdot\tau_{\rm BLR}$, e.g. \citealt{2017ApJ...851...21G}; \citealt{2021ApJ...915..129K}), and between the AD continuum and the emission of a putative dust torus (or torus size; $R_{\rm dust} \propto c\cdot\tau_{\rm dust}$, e.g., \citealt{2019MNRAS.489.1572L}; \citealt{2020ApJ...891...26A}).
By combining spectroscopic and photometric observations, the method has revealed black hole masses ($M_{\rm BH}$), and Eddington ratios in hundreds of AGN (see \citealt{2021iSci...24j2557C} and references therein). 

Successful recovery of the time delay from an RM campaign depends on observational noise, intrinsic variability, and how well the light curves are sampled with respect to the delay. Highly sampled RM light curves may constrain the geometry of the reverberant region (e.g., \citealt{2012ApJ...754...49P}; \citealt{2013ApJ...764...47G}; \citealt{2014A&A...568A..36P}).
Astronomical observations, however, are often affected by weather conditions, technical problems, sky coverage in the case of satellites, and seasonal gaps.
Some of those effects are even more severe when observing high-redshift quasars.
In this context, several algorithms have been proposed to deal with irregular time sampling.
For example, traditional cross-correlation techniques applied to linearly interpolated data (ICCF; \citealt{1987ApJS...65....1G}; \citealt{1999PASP..111.1347W}), the discrete correlation function (DCF; \citealt{1988ApJ...333..646E}, including the Z-transformed DCF of \citealt{1997ASSL..218..163A}), statistical models characterised by stochastic processes (\citealt{1992ApJ...398..169R}) as implemented in the code JAVELIN (\citealt{2011ApJ...735...80Z}), with the assumption of a damped random walk model for AGN variability (\citealt{2009ApJ...698..895K}, \citealt{2010ApJ...708..927K}) and the Von Neumann estimator (\citealt{2017ApJ...844..146C}), which does not rely on the interpolation and binning of the light curves, but on the degree of randomness of the data. 

An important limitation affecting all the above methods is that the results change significantly with sparser time sampling and large flux uncertainties. Furthermore, they do not lend themselves to model comparison as they do not provide predictions on test data (e.g. held-out data in a cross-validation scheme).

Gaussian processes (GP) have been used to model the AGN light curves to mitigate the effects of gaps and improve the estimation of the time delay (see \citealt{2016ApJ...819..122Z} and the implementation in JAVELIN). The code JAVELIN assumes a model where the time-shifted signal is the result of the convolution between the driver and a top-hat transfer function.

A less model-dependent approach that does not rely on the shape of the transfer function is provided by the ICCF (\citealt{1987ApJS...65....1G}), which is still the most widely used method for estimating time delays in current RM research (see \citealt{2022arXiv221209161P} and references therein). However, the ICCF is very sensitive to the input chosen parameters. The search range of the time delay, the size of the interpolation, whether it is the peak or the centroid of the ICCF, the threshold used to calculate these two quantities and other assumptions (\citealt{1999PASP..111.1347W}) can lead to very different estimates.

In this work, we seek to address the above issues and propose a model that reformulates the ICCF in a probabilistically sound fashion.
The proposed model is based on a Gaussian Process, and we name it the Gaussian Process Cross-Correlation (GPCC) model. 
The paper is organized as follows: after introducing relevant notation, we briefly describe certain shortcomings of the ICCF method. We then describe the proposed GPCC model in Section~\ref{sec2}. In Section~\ref{sec3} we demonstrate the behaviour of the GPCC model on simulated data. In Section~\ref{sec4} we apply the method to real observations.
We finally draw our conclusions and summarise the main results in Section~\ref{sec5}.

\section{Methods}\label{sec2}

After introducing notation, we briefly review the ICCF and discuss its shortcomings and modelling assumptions. Based on these modelling assumptions, we present a probabilistic
reformulation of ICCF.

\subsection{Notation for light curves}
\label{sec:notation}

Observed data are composed of $L$ number of  light curves $y_l(t)$, each observed at one of the $L$ number of bands.
Light curve $y_l(t)$ is observed at $N_l$ number of observation times $\bd{t}_l = [t_{l,1}, \dots, t_{l,N_l}] \in\RR{N_l}$.

We denote the flux measurements of the $l$-th  light curve at these times as 
$\bd{y}_l = [y_l(t_{l,1}), \dots, y_l(t_{l,N_l})]\in\RR{N_l}$. 
Also, we denote the errors of the flux measurements of the $l$-th light curve as
$\bd{\sigma}_l = [\sigma_{l}(t_{l,1}), \dots, \sigma_{l}(t_{l,N_l})]\in\RR{N_l}$.
We define the total number of measurements in the dataset as $N=\sum_{l=1}^L N_l$.
Notation $\bd{y} = [\bd{y}_1, \dots, \bd{y}_L] \in \RR{N}$ stands for concatenating
the flux measurements of all bands. 
Similarly, we also use $\bd{t} = [\bd{t}_1, \dots, \bd{t}_L] \in \RR{N}$ and $\bd{\sigma} = [\bd{\sigma}_1, \dots, \bd{\sigma}_L] \in \RR{N}$.

We associate the $l$-th light curve with a scale parameter $\alpha_l>0$, an offset parameter $b_l$ and a delay parameter $\tau_l$.
We collectively denote these parameters as vectors $\bd{\alpha}=[\alpha_1,\dots,\alpha_L]$, $\bd{b}=[b_1,\dots,b_L]$ and $\bd{\tau}=[\tau_1,\dots,\tau_L]$.

Throughout this work, whenever we concatenate  data or parameters pertaining to the $L$ light curves, we always do it in the order of $1$-st to $L$-th band, as demonstrated by e.g.~$\bd{y} = [\bd{y}_1, \dots, \bd{y}_L]$.

\subsection{Interpolated cross-correlation function}\label{iccfsec}

For simplicity, we review the ICCF for the case of two signals $y_{\rm conti}(t)$ and $y_{\rm line}(t)$, which in RM applications often correspond to light curves for the AD continuum and the line emission from the BLR, respectively.
Following \cite{1987ApJS...65....1G}, the correlation function $\rm{CCF}(\tau)$ between the two signals for a time delay of $\tau$ reads:
\begin{eqnarray}
 \frac{\EX \big\{ \big[ y_{\rm conti}(t) - \EX[y_{\rm conti}(t)] \big] \big[ y_{\rm line}(t+\tau) - \EX[y_{\rm line}(t)] \big]   \big\}}{\sigma_{\rm conti} \sigma_{\rm line}} \ ,
\end{eqnarray}

where  $\EX$ is expectation, and $\sigma_{\rm conti}$, $\sigma_{\rm line}$ are the standard deviations of the two light curves. 

The ICCF proceeds as follows:
First it shifts the continuum curve $y_{\rm conti}(t)$ by $t + \tau$ and the line emission $y_{\rm line}(t)$ is linearly interpolated at times $t + \tau$ that are matching the time range between the minimum and maximum of the observed line curve.
Points outside this range are excluded from the calculations.
Then, the CCF is calculated for this two new time series.
In other words, we shift the continuum $y_{\rm conti}(t)$ by the time delay $\tau$ and compute the CCF between the shifted continuum and the interpolated line curve $y_{\rm line}(t)$, thus providing the ICCF value for the interpolated line emission light curve, which we denote as $\rm ICCF_{\rm line, \tau}$.
Next, the process is repeated, but this time the continuum light curve is linearly interpolated and then correlated with the observed line emissions at time $t-\tau$, hereby retrieving $\rm ICCF_{\rm conti, \tau}$.
The final interpolated cross correlation function value is given by the average,
\begin{eqnarray}
\rm{ICCF}(\tau) = \frac{ {\rm ICCF_{\rm line,\tau}} + \rm ICCF_{\rm conti,\tau}}{2}
\label{eq:eqccf}
\end{eqnarray}

and has to be calculated for each respective delay $\tau$.

The time delay is given by the centroid on Equation~\eqref{eq:eqccf}, which is calculated for values above 80\% of the $\rm{ICCF}(\tau)$ peak value.
While this lower limit has been widely used in RM studies, well-sampled data allow the use of even lower values, down to 50\% (e.g., \citealt{2012A&A...545A..84P, 2015A&A...576A..73P}; see also Appendix in \citealt{2004ApJ...613..682P}).
In the appendix in Figure \ref{fig:iccfexamp} we show an example of the application of ICCF to simulated data for the AD continuum and the
BLR emission light curves.

A popular method to provide uncertainty estimates for derived delays in ICCF is the bootstrapping method, or better known in the RM field as \emph{the flux randomization and random subset selection} method (FR/RSS, \citealt{1998PASP..110..660P}, 
 \citealt{2004ApJ...613..682P}).
The FR/RSS method works as follows:
A subset (typically $\sim2000$ light curves) is randomly generated from the observed light curves, with each new light curve containing only 63\% of the original data points\footnote{Considering the Poisson probability that no particular point is selected, the size of the selected sample is reduced by a factor of about 1/e, resulting in 63\% of the original data (see \citealt{2004ApJ...613..682P})}.
The flux value of each data point is randomly perturbed according to the assumed normally distributed measurement error.
The ICCF is then calculated for the pairs of subset light curves, resulting in a centroid (or peak) distribution from which the uncertainties are estimated from the 68\% confidence interval (Figure~\ref{fig:iccfexamp}).

\subsection{Shortcomings with the ICCF}

In the following, we review certain issues when using the ICCF for determining the delay.

\subsubsection{Dealing with more than two light curves}
\label{sec:morethantwolightcurves}

The ICCF considers only pairs of light curves at a time.
In order to estimate delays between more than two light curves, ICCF chooses one of the light curves as a reference.
Delays are then estimated with respect to the reference light curve and every other light curve in the dataset. 
Hence, instead of estimating the delays in a joint manner, the problem is broken into multiple independent delay estimation problems.
This discards the fact that the delay between one pair of light curves may affect the delay between another.

\subsubsection{80\% rule, peak and centroid}
In the context of RM of the BLR, the peak of the cross-correlation provides an estimate of the inner size of the BLR, i.e. the response of the gas located at small radii. 
This can be understood as a bias of the cross-correlation for cases where the BLR is extended in radius (e.g. spherical or disc-shaped geometries). 
The centroid of the cross-correlation, on the other hand, gives the luminosity-weighted radius and is mathematically equivalent to the centroid of the transfer function (see derivation in \citealt{1991ApJS...75..719K}). 
The ICCF centroid is therefore an important quantity for which there is no simple calculation method. 
The reason is that it depends considerably on the quality of the light curves, i.e. on the noise and the sampling. 
For a noislees light curve with ideal sampling, the ICCF peak is well defined and the centroid estimation is straightforward. 
However, with noisy and unevenly sampled data, the ICCF may result in multiple peaks. 
In this case, the centroid is calculated using the ICCF values around the most significant peak. 
With multiple peaks, this is obviously a major challenge.

\begin{figure}
    \centering
    \includegraphics[width=\columnwidth]{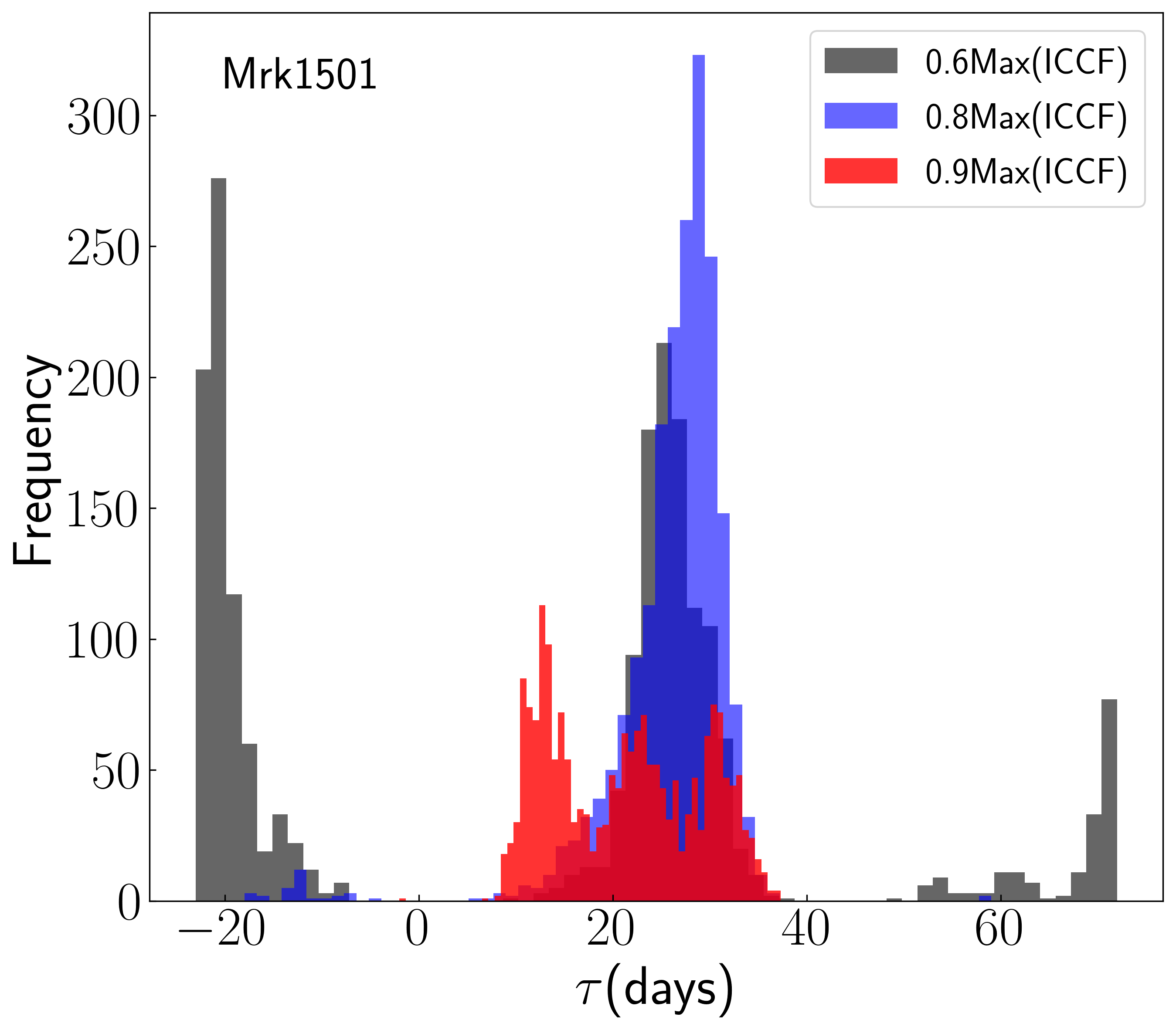}
    \caption{The histograms show the distribution of the centroid time delay for Mrk1501 obtained with the FR/RSS method for thresholds 0.6 (black) 0.8 (blue), and 0.9 (red).}
    \label{fig:frssmrk}
\end{figure}

Using Monte Carlo simulations, \cite{2004ApJ...613..682P} has suggested that a threshold of 80\% of the $\rm{ICCF}(\tau)$ peak is a good compromise based on the width of the obtained centroid distributions. 
Lower values for the threshold are also conceivable if the peak of the ICCF is too noisy. 
Consequently, the calculation of the threshold value for the centroid must be decided on a case-by-case basis and is difficult to standardise, especially when the ICCF is applied to a large number of objects. To illustrate the effects of threshold selection, we show in Figure~\ref{fig:frssmrk} the distributions of delays obtained with the FR/RSS method for three thresholds (0.6, 0.8 and 0.9). For this example, we have chosen an object where the ICCF appears broad and without a clear peak (Figure~\ref{fig:GPCCexampMrk1501}), which makes the results particularly sensitive to the choice of thresholds.

\subsubsection{Out-of-sample performance}

One way to test how well a model generalizes to future data is to evaluate its predictive performance on data that do not belong to training sample, such as test data. If the model predicts well on out-of-sample data, it means that it has successfully captured certain aspects of the process underlying the observed data, i.e. the model can generalize.
The standard ICCF (as described in Section~\ref{iccfsec}) does not provide predictions for out-of-sample data that would allow us to assess its generalisation performance and compare against alternative, competing models.

\subsubsection{Absence of delay prior}

The ICCF is often used together with the FR/RSS method to obtain a probability distribution for the delay. However, the FR/RSS method does not consider a prior distribution for the delay. The proposed GPCC considers a prior distribution for the delay. In particular, we give an example of a prior based on BLR photoionisation physics (see section \ref{sec:prior_on_delay}), which has the property of suppressing long delays that seem implausible. Previously, the alias mitigation technique was used to suppress long delays that occur due to a small number of overlapping points (e.g. \citealt{2021ApJ...912...10Z}). However, alias mitigation suffers from the following issues: a) it is not part of the respective probabilistic model formulations and yet it contributes to the calculation of the posterior, b) it is not physically motivated, and c) it cannot be interpreted as a prior because it depends on the data. Although our choice of prior may seem subjective, our method is not dependent on this particular choice and can indeed consider alternative priors. However, discussing which physical prior is more appropriate is a more fruitful approach than devising weighing schemes based on heuristic motivation.

\subsection{Reformulation of cross-correlation method}

We examine the main modelling assumption of the ICCF.
We take this assumption on board and propose a probabilistic reformulation.

\subsubsection{Assumptions of ICCF}
\label{sec:implicit_assumption}

In this section, we present the modelling assumptions underlying the ICCF method.
The cross-correlation function between two time series  $y_1(t)$ and $y_2(t)$ is a measure of their overlap and reads:
\begin{equation}
(y_1\star y_2)(\tau) =\int_{-\infty}^\infty
y_1(t) y_2(t+\tau) dt \ ,
\end{equation}
where $\tau$ is denoted as the delay. 

We consider two cases in which time series may be related.
In the first case, we assume that the time series are related by $y_2(t-\tau) = \alpha y_1(t)+b$.
That is, $y_2$ is a scaled, offset, and delayed version of $y_1$.
If $\tau$ is unknown, we can estimate it as $\hat{\tau} = \argmax_{\tau}(y_1\star y_2)(\tau)$.
In this case, the maximum overlap between the two time series $y_1(t)$ and $y_2(t)$ at $\hat{\tau}$ is an estimate of the lag which aligns the two time series.

In the second case, we consider a more general relation between two time series, $y_2 = \mathcal{T}\{ y_1\}$,  e.g.~$\mathcal{T}$ may stand for convolution with a function $h$ such that $y_2(t) = \int_{-\infty}^{\infty} h(\tau) y_1(t-\tau) d\tau$.
In this case, $y_2(t)$ may have certain features (e.g., peaks or troughs) that also occur in $y_1(t)$ but at an earlier time.
Therefore, $y_2(t)$ may look like a delayed version of $y_1(t)$ and we can  use 
 again $\argmax_{\tau}(y_1\star y_2)(\tau)$ to estimate this perceived lag between the respective features of the light curves.
Of course, in this second case we cannot find $\tau$ such that a delayed time series $y_2(t-\tau)$ aligns with $y_1(t)$, since $y_2$ is not simply a time-shifted version of $y_1$ but a transformation of it described by $\mathcal{T}$.

We argue that ICCF relies on the assumption that the light curves are related by $\mbox{$y_2(t-\tau) = \alpha y_1(t)+b$}$, since this is the only case in which the maximum overlap coincides with an estimate of the delay between two time series.
We also note that two time series related by $\mbox{$y_2(t-\tau) = \alpha y_1(t)+b$}$, with respect to a latent signal $f(t)$ can be written as follows:
\begin{eqnarray}
y_1(t) &=& \alpha_1 f(t-\tau_1)+b_1 \ , \notag\\
y_2(t) &=& \alpha_2 f(t-\tau_2)+b_2 \ ,
\end{eqnarray}
where each time series $y_l(t)$ has its own scale $a_l$, offset $b_l$ and delay $t_l$.  

Introducing a common latent signal $f(t)$, which is the unobserved driver of the observed time series, allows us to indirectly relate more than just two time series:
\begin{eqnarray}
y_1(t) &=& \alpha_1 f(t-\tau_1)+b_1 \notag\\
\vdots \notag\\
y_L(t) &=& \alpha_L f(t-\tau_L)+b_L \ .
\end{eqnarray}
These relations allow us to consider the joint estimation of multiple delays $\tau_1, \dots, \tau_L$ between $L$ number of time series.
Hence, when we use cross-correlation to reveal the delay between multiple light curves, we implicitly assume that they are related via a latent signal:
\begin{equation}
y_l(t) = \alpha_l f(t-\tau_l)+b_l \ .
\label{eq:relation_via_latent}
\end{equation}

\subsection{Gaussian Process Cross-Correlation}

In this section, we propose to model $f(t)$ as a Gaussian process (GP).
This leads to a model that we refer to as {\it Gaussian Process Cross-Correlation} (GPCC).

\subsubsection{Model formulation}
\label{sec:model_formulation}

Based on  Equation~\eqref{eq:relation_via_latent}, we assume that  a common latent signal $f(t)$ generates all observed light curves.
Here we postulate that $f(t)\sim\mathcal{GP}(0,k_\rho)$, meaning that $f(t)$ is drawn from a zero-mean GP with covariance function $k_\rho(t,t^\prime)$, where $\rho$ is a scalar parameter.
For two observed light curves $y_i(t)$ and $y_j(t)$ we write:
\begin{align}
y_i(t_{i,n}) &= \alpha_i f(t_{i,n}- \tau_i)+ b_i+ \epsilon_{i,n} \ , \notag\\
y_j(t_{j,m}) &= \alpha_j f(t_{j,m}- \tau_j)+ b_j+ \epsilon_{j,m} \ ,
\label{eq:generative_model}
\end{align}
where $\epsilon_{i,n}$ is zero-mean Gaussian noise with standard deviation $\sigma_{i,n}$.
We impose a Gaussian prior on the offset vector
$p(\bd{b})=\mathcal{N}(\bd{b}|\bd{\mu}_b,\bd{\Sigma}_b)=\prod_{l=1}^L
\mathcal{N}(b_l|\mu_{b_l},\sigma_{b_l}^2)$.
Given that $f(t)$ is drawn from a GP and that a GP is closed under affine transformations, the joint distribution of the observed light curves is also governed by a GP. 
To specify this GP, we need its mean and covariance function.
Taking expectations over both priors $\mathcal{GP}(0,k_\rho)$ and $p(\bd{b})$, we calculate the mean for the $i$-th band:
\begin{align}
&  \mu_{b_i} = \EX \big[y_i(t_{i,n}) \big] \ .
\label{eq:mean_function}
\end{align}
Taking again expectations over both priors, we calculate the covariance between the fluxes observed in the $i$-th band and $j$-th band\footnote{The Kronecker delta $\delta_{ij}$ equals $1$ when $i=j$, otherwise $0$.}:
\begin{align}
%
c(t_{i,n}, t_{j,m}) =&
 \EX \big[y_i(t_{i,n}) - \mu_{b_i} \big] \EX \big[y_j(t_{j,m})) - \mu_{b_j} \big] \notag\\
=&\ \alpha_i\alpha_j k_\rho(t_{i,n}-\tau_i, t_{j,m}-\tau_j) \notag\\
 & + \delta_{i,j} \sigma_{b_i}^2  + \delta_{i,j} \delta_{n,m} \sigma_{i,n}^2
\
\label{eq:cov_function}
\end{align}
Evaluating $c(\cdot,\cdot)$ at all possible pairs we can form by pairing the observations times $t_{i,n}$ in the $i$-th band with $t_{j,m}$ in $j$-th band, results in a $N_i\times N_j$ covariance matrix $\bd{C}(\bd{t}_i,\bd{t}_j)$.
We repeat this calculation for each of the $(i,j)$ pairs of bands to obtain $L^2$ number of covariance matrices.
We arrange these individual covariance matrices in a $L\times L$ block structure to form the $N\times N$ covariance matrix $\bd{C}(\bd{t},\bd{t})$ between all possible pairs of bands.
For instance, for $L=2$ bands we would have:
\begin{equation}
\bd{C}(\bd{t},\bd{t}) = 
    \begin{pmatrix}
    \bd{C}(\bd{t}_1,\bd{t}_1) & \bd{C}(\bd{t}_1,\bd{t}_2) \\
    \bd{C}(\bd{t}_2,\bd{t}_1) & \bd{C}(\bd{t}_2,\bd{t}_2) \ 
    \end{pmatrix} \ .
    \label{eq:cov_function_example}
\end{equation}
Hence, the $(i,j)$-th block of $\bd{C}(\bd{t},\bd{t})$ has dimensions $N_i\times N_j$ and holds the covariances between the $i$-th and $j$-th light curves; the $(n,m)$-th entry of the $(i,j)$-th block is given by $c(t_{i,n}, t_{j,m})$.

Given the derived mean and covariance, the  likelihood reads:
\begin{equation}
    p(\bd{y}|\bd{t}, \bd{\sigma}, \bd{\tau}, \bd{\alpha}, \rho) = \mathcal{N}(\bd{y}|\bd{Q} \bd{\mu}_b, \bd{C}(\bd{t},\bd{t})) \ .
    \label{eq:gp_likelihood}
\end{equation}
\bd{Q} is an auxiliary $N\times L$ matrix, with entries set to $0$ or $1$ (see appendix \ref{app:matrix_q}), that replicates the vector $\bd{\mu}_b$ so that: $$\bd{Q}\bd{\mu}_b=[\underbrace{\mu_{b_1}, \dots \mu_{b_1}}_{N_1},\dots,\underbrace{\mu_{b_L}, \dots, \mu_{b_L}}_{N_L}]\ . $$

\noindent
Henceforth, we suppress in the notation the conditioning on $\bd{t}$ and $\bd{\sigma}$ and consider it implicit, i.e. $p(\bd{y}|\bd{t}, \bd{\sigma}, \bd{\tau}, \bd{\alpha},\rho) =
p(\bd{y}|\bd{\tau},\bd{\alpha},\rho)$.
Finally, fitting the observed data  involves maximizing the likelihood $p(\bd{y}|\bd{\tau},\bd{\alpha},\rho)$ with respect to the free parameters $\bd{\alpha}$ and $\rho$.

\subsubsection{Prior on delay}
\label{sec:prior_on_delay}

Photoionization physics define the ionization parameter as
\begin{equation}
U = \frac{Q(\rm{H})}{4\pi r^2 c\, \eta_{\rm H}} \,
\end{equation}
where $Q(\rm{H})$ is the number of photons per second emitted from the central source ionizing the hydrogen cloud, $r$ is the distance between the central source and the inner face of the cloud, and $\eta_{\rm H}$ is the total hydrogen density.
Assuming that the BLR's for low luminosity Seyfert to high-luminosity quasars have the same ionization parameter and BLR density, one can define an upper limit on the BLR size as a function of the AGN bolometric luminosity,  
\begin{equation}
R_{\rm{BLR}}\propto Q(\rm{H})^{1/2} \propto L_{\rm{bol}}^{1/2} \ .
\end{equation}
Since $L_{\rm {bol}}$ is in practice very difficult to measure, one can use bolometric corrections, e.g., using the optical continuum luminosity measured at $5100$\AA,
$L_{\rm Bol} = 10\lambda L_{\lambda}(5100$\AA) (\citealt{2004MNRAS.352.1390M}). 
\cite{2013ApJ...767..149B} provides a normalized expression for the H$\beta$ BLR size for a given optical continuum luminosity
\begin{equation}
R_{\rm {H}\beta}(\lambda L_{\lambda}(5100\AA))= 10^{1.559} \bigl[ \lambda L_{\lambda}(5100\AA)/10^{44} \bigr]^{0.549} [light-day] \ .
\end{equation}
where the slope value of 0.549 corresponds to the Clean2+ExtCorr fit obtained from \cite{2013ApJ...767..149B} (their Table 14), which includes a special treatment of the sources and an additional extinction correction.

We define $\tau_{max}(l, z)=R_{\rm {H}\beta}(l) (1+z) [light-day]$ as the upper limit on the delay measured in the observer frame, where we define $l = 10\lambda L_{\lambda}(5100\AA)$ to simplify notation.
Since we have no further information, we assume the prior to be the uniform distribution:
\begin{equation}
p(\tau)=\mathcal{U}(0,\tau_{max}(l,z)) \ .
\end{equation}
The uniform distribution is the maximum entropy probability distribution for a random variable about which the only known fact is its support.

\subsubsection{Predictive likelihood for GPCC}

We wish to compute the likelihood on new light curve data, i.e.~test data, as opposed to training data on which our model is already conditioned.
Following the notation in Section~\ref{sec:notation}, we denote test data with an asterisk such that $\bd{y}^*=[\bd{y}^*_1,\dots,\bd{y}^*_L]$, where each light curve $\bd{y}^*_l= [y^*(t_{l,1}), \dots, y^*(t_{l,N_l})]\in\RR{N^*_l}$ is observed at times $\bd{t}^*_l = [t^*_{l,1}, \dots, t^*_{l,N^*_l}] \in\RR{N^*_l}$ with measured errors $\bd{\sigma}^*_l = [\sigma^*_{l,1}, \dots, \sigma^*_{l,N^*_l}] \in\RR{N^*_l}$.
Accordingly, we define $N^*=\sum_{l=1}^L N^*_l$.
We also define the cross-covariance function between training and test data
\begin{align}
& c^*(t^{~}_{i,n}, t^*_{j,m}) = \alpha_i\alpha_j k_\rho(t^{~}_{i,n}-\tau_i, t^*_{j,m}-\tau_j)+ \delta_{i,j} \sigma_{b_i}^2 
\ ,
\label{eq:crosscov_function}
\end{align}
which is identical to Equation \eqref{eq:cov_function} after discarding its last term.
Evaluating $c^*(\cdot,\cdot)$ at all possible pairs we can form by pairing the observations times $t_{i,n}$ in the $i$-th band with the test observation times $t^*_{j,m}$ in $j$-th band, results in a $N_i\times N^*_j$ covariance matrix $\bd{C}^*(\bd{t}_i,\bd{t}^*_j)$.
By repeating this for all $(i,j)$ pairs of bands, we obtain $L^2$ number of covariance matrices, which we arrange in a $L\times L$ block structure to form the $N\times N^*$ covariance matrix $\bd{C}^*(\bd{t},\bd{t}^*)$.
For example, in the case of $L=2$ bands we would have:
\begin{equation}
\bd{C}^*(\bd{t},\bd{t}^*) = 
    \begin{pmatrix}
    \bd{C}^*(\bd{t}^{~}_1,\bd{t}^*_1) & \bd{C}^*(\bd{t}^{~}_1,\bd{t}^*_2) \\
    \bd{C}^*(\bd{t}^{~}_2,\bd{t}^*_1) & \bd{C}^*(\bd{t}^{~}_2,\bd{t}^*_2)
    \end{pmatrix} \ .
\end{equation}
The predictive likelihood for the new data (test data) given the observed data (training data) is a Gaussian distribution\footnote{We suppress in the notation the conditioning on $\bd{t},\bd{\sigma},\bd{t}^*,\bd{\sigma}^*$.} (see Appendix \ref{app:predictive_likelihood}): 
\begin{equation}
p(\bd{y}^*|\bd{y},\bd{\tau},
\bd{\alpha}, \rho) = \mathcal{N}(\bd{y}^*|\bd{\mu}^*, \bd{\Sigma}^*)    \ ,
\label{eq:predictive_likelihood}
\end{equation}
where
\begin{align}
&\bd{\mu}^* = \bd{Q}^*\bd{\mu}_b+\bd{C}^*(\bd{t},\bd{t}^*)^T\bd{C}(\bd{t},\bd{t})^{-1}(\bd{y}-\bd{Q}\bd{\mu}_b) \ , \\
&\bd{\Sigma}^* = \bd{C}(\bd{t}^*,\bd{t}^*) - \bd{C}^*(\bd{t},\bd{t}^*)^T \bd{C}(\bd{t},\bd{t})^{-1} \bd{C}^*(\bd{t},\bd{t}^*) \ .
\end{align}
The $N^*\times L$ auxiliary matrix $\bd{Q}^*$ replicates vector $\bd{b}$ so that:
$$\bd{Q}^*\bd{\mu}_b=[\underbrace{\mu_{b_1}, \dots, \mu_{b_1}}_{N^*_1},\dots,\underbrace{\mu_{b_L}, \dots, \mu_{b_L}}_{N^*_L}]\ .$$
We construct the matrix $\bd{Q}^*$ in the same way as the matrix \bd{Q}, as shown in the appendix \ref{app:matrix_q}.
The $N^*\times N^*$ matrix $\bd{C}(\bd{t}^*,\bd{t}^*)$ is computed in the same way as the matrix $\bd{C}(\bd{t},\bd{t})$, i.e., by evaluating $c(\cdot,\cdot)$ at all possible pairs $t^*_{i,n}, t^*_{j,m}$.
For example, in the case of $L=2$ bands we would have:
\begin{equation}
\bd{C}(\bd{t}^*,\bd{t}^*) = 
    \begin{pmatrix}
    \bd{C}(\bd{t}^*_1,\bd{t}^*_1) & \bd{C}(\bd{t}^*_1,\bd{t}^*_2) \\
    \bd{C}(\bd{t}^*_2,\bd{t}^*_1) & \bd{C}(\bd{t}^*_2,\bd{t}^*_2)
    \end{pmatrix} \ .
\end{equation}

\subsubsection{Posterior density for delays}

Given a data set $\bd{y}, \bd{t}, \bd{\sigma}$ observed at $L$ number of bands, we want to infer the posterior distribution for $\bd{\tau}$. 
We compute the non-normalized posterior of the delays over a finite regular grid of delay combinations, and then normalize these values into a multinomial distribution.
We denote this grid and the multinomial posterior probabilities on the grid by
\begin{align}
    \mathcal{T} &= \{\bd{\tau}_1,\dots,\bd{\tau}_T\} , \notag \\ 
    \pi_i &= \frac{p(\bd{\tau}_i|\bd{y}, \hat{\bd{\alpha}}_i, \hat{\rho}_i)}{\sum_{j=1}^T p(\bd{\tau}_j|\bd{y},  \hat{\bd{\alpha}}_j, \hat{\rho}_j)}
     = \frac{p(\bd{y}|\bd{\tau}_i, \hat{\bd{\alpha}}_i, \hat{\rho}_i)p(\bd{\tau}_i)}{\sum_{j=1}^T p(\bd{y}|\bd{\tau}_j, \hat{\bd{\alpha}}_j, \hat{\rho}_j)p(\bd{\tau}_j)}
\label{eq:posterior}
\end{align}
with elements $\bd{\tau}_i$ in the set $\{0,1\delta\tau,2\delta\tau,\dots, T\delta\tau\}^L$.
Here $\hat{\bd{\alpha}}_i$ and $\hat{\rho}_i$ are the optimal parameters obtained when maximizing the likelihood in \eqref{eq:gp_likelihood} for each considered delay $\bd{\tau}_i$.
Since delays are relative to each other, e.g.~delays $\tau_1 = 0.3, \tau_2 = 1.3$ are equivalent to $\tau_1 = 0.0, \tau_2 = 1.0$, without loss of generality we fix the delay parameter of the first light curve to $\tau_1=0$.

\section{Simulations}\label{sec3}

\begin{figure*}
    \centering
    \includegraphics[width=\textwidth]{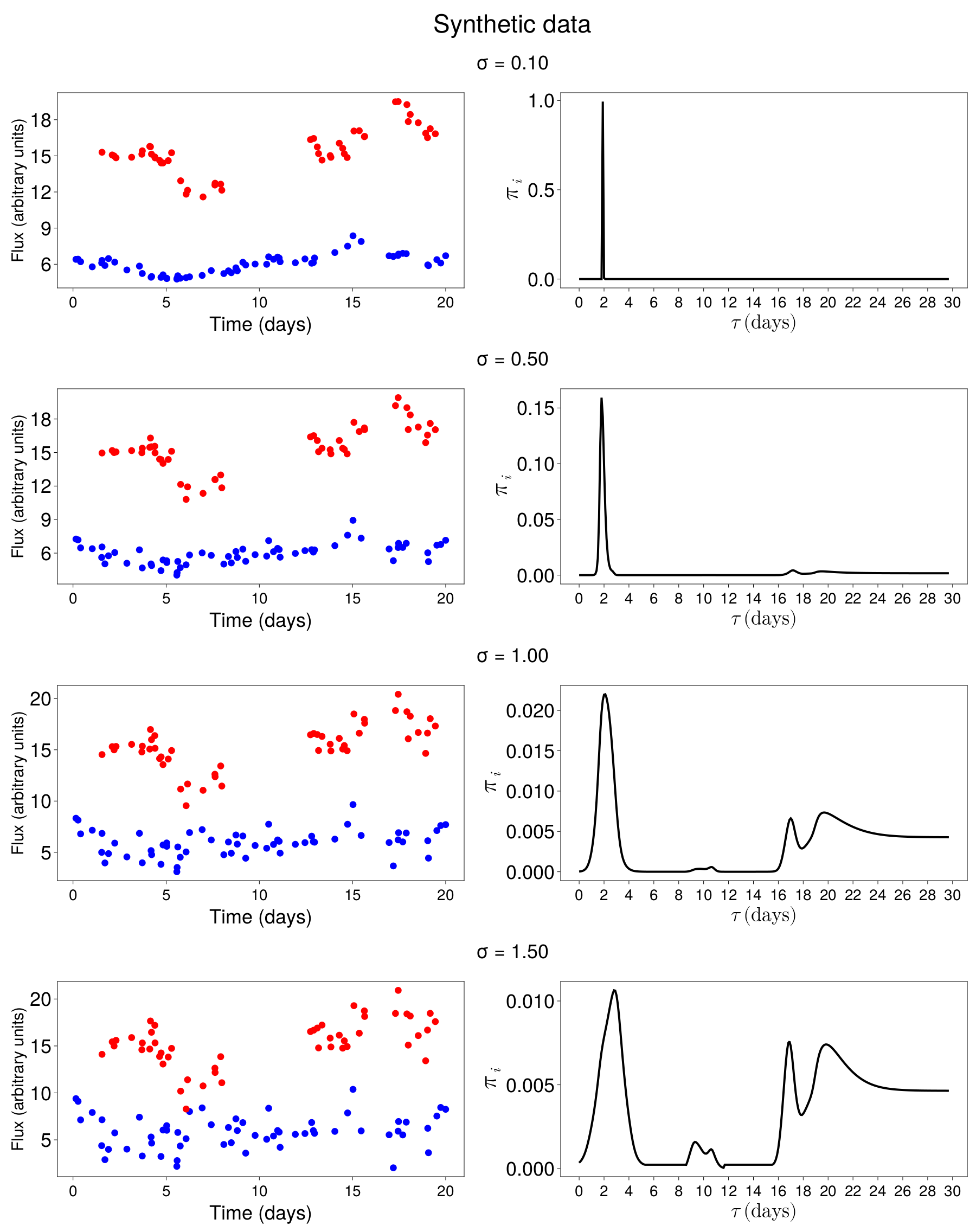}
    \caption{Synthetic simulations with true delay at $2$ days.
    The vertical axis is in arbitrary units of simulated flux. 
    The left column shows the simulated light curves.
    The inferred posterior distribution of delay is shown in the right column.
    The noise $\sigma$ increases from the top to the bottom row.}
    \label{fig:synthetic_simulation}
\end{figure*}

We present a simple numerical simulation with synthetic data that shows how the model behaves under perfect theoretical conditions, i.e. when the data match our model assumptions.

We simulate data in two bands with a delay of $\tau=2$ days, i.e.~$\tau_1=0, \tau_2=2$.
The remaining parameters are set as follows: $\alpha_1=1$, $\alpha_2=1.5$, $b_1=6$, $b_2=15$.
We use the Ornstein-Uhlenbeck (OU) kernel $k_\rho(t,t^\prime) = \exp(-\frac{\|t-t^\prime \|}{\rho})$ and set $\rho=3.5$.
For the first band, we sample $60$ observation times from the uniform distribution $\mathcal{U}(0,20)$.
For the second band, we sample $50$ observation times from the two-component distribution $0.5\mathcal{U}(0,8)+0.5\mathcal{U}(12,20)$, so that an observation gap is introduced between days $8$ and $12$.
After sampling the observation times, we sample the latent signal $f(t)\sim\mathcal{ GP }(0,k_\rho)$ and then the fluxes using Equation~\eqref{eq:generative_model}.
We perform four experiments in which we attempt to recover the delay under Gaussian noise using the respective standard deviations
$\sigma\in\{0.1, 0.5, 1.0, 1.5\}$.
Since delays are relative to each other, e.g.~delays $\tau_1 = 0.3, \tau_2 = 1.3$ are equivalent to $\tau_1 = 0.0, \tau_2 = 1.0$, we fix $\tau_1 = 0.0$ and seek only the second delay $\tau=\tau_2$. We assume the prior $p(\tau)\propto 1$ and consider candidates $\tau$ in the grid $\{0,0.1,\dots,30\}$ of step size $0.1$ days.

\begin{figure}
    \centering
    \includegraphics[width=\columnwidth]{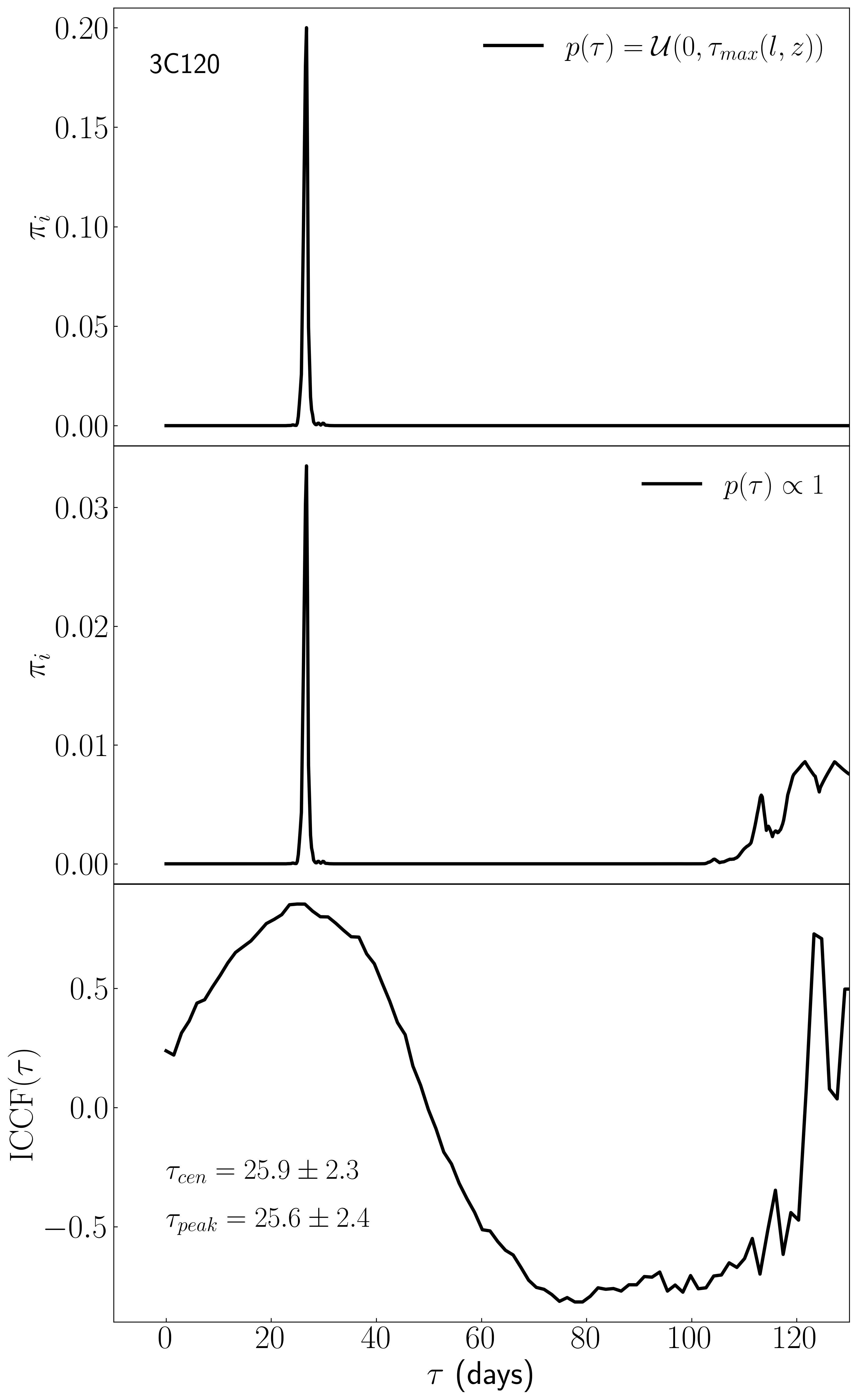}
    \caption{GPCC application to 3C120. Top: posterior distribution of delay with prior $p(\tau)=\mathcal{U}(0,\tau_{max}(l,z))$.
    The peak with the highest probability corresponds to a delay of 27.6 days.
    Middle: posterior distribution of delay with flat prior $p(\tau)\propto 1$.
    Bottom: ICCF calculated with a sampling of 1 day for linear interpolation of fluxes. The centroid is determined using the $80\%$ rule.
    Our results for the ICCF peak and centroid are identical to those of G+2012, and we show the results of G+2012 including the uncertainty estimated with the FR/RSS method.
    All delays are indicated in the rest-frame.}
    \label{fig:GPCCexamp}
\end{figure}

We show the results of the simulations in Figure~\ref{fig:synthetic_simulation}.
The true peak at $2$ days is identified in all experiments.
For the values of $\sigma=1.0$ and $\sigma=1.5$ we see that
alternative peaks arise as potential candidates. The reason that new peaks arise is because noise can suppress salient features, making light curves look more similar to each other.
We note in particular that alternative peaks tend to arise at large delays, which we explain as follows. Small delays shift light curves only slightly so that they still overlap considerably after the shift (see left plot in Figure~\ref{fig:GPCCexampsimul}). For a small delay to be likely, the algorithm must match a number of features (i.e. peaks, troughs) appearing in one light curve to features appearing in the other. Large delays, however, shift light curves by so much that the they overlap only slightly after the shift (see right plot in Figure~\ref{fig:GPCCexampsimul}). For a large delay to be likely, the algorithm must match fewer light curve features than in the case of a small delay. Hence, it is "easier" for a large delay to arise as a candidate than it is for a small delay\footnote{Alias mitigation (e.g. ~\citealt{2017ApJ...851...21G, 2021ApJ...912...10Z}) has been used previously to address the situation where a small number of overlapping points leads to peaks in the distribution of delays.}

We also observe that the posterior distribution becomes flat in the interval of $[20,30]$ days. Recall that the simulated light curves span no longer than $20$ days. A delay in $[20,30]$ shifts the light curves so far apart that they no longer overlap\footnote{Unlike the ICCF, such a delay is allowed in our formulation and the model assigns  a probability to it.} (see right plot in  Figure~\ref{fig:GPCCexampsimul}). 
All large delays that lead to non-overlapping light curves yield the same covariance matrix (see Equation~\ref{eq:cov_function_example}) 
\begin{equation}
\bd{C}(\bd{t},\bd{t}) = 
    \begin{pmatrix}
    \bd{C}(\bd{t}_1,\bd{t}_1) & \bd{0} \\
    \bd{0} & \bd{C}(\bd{t}_2,\bd{t}_2) \ 
    \end{pmatrix} \ \mbox{where} \ \bd{C}(\bd{t}_1,\bd{t}_2) = \bd{0}
\end{equation}
and thus the same likelihood, which explains why the posterior distribution becomes and stays flat after a certain delay.

\section{Data applications}\label{sec4}

In this section, we apply our method to AGN data from RM campaigns for which time delay measurements were made using the ICCF method.
In particular, we focus on the sample of five AGN from \cite{2012ApJ...755...60G} (G+2012) that provided high sampling rate light curves in the continuum (at 5100\AA) and in the H$\beta$ emission line as part of a BLR RM campaign. For an application of the method to more than two light curves, we use the data obtained for MCG+08-11-011 as part of a RM monitoring of the accretion disc by \cite{2018ApJ...854..107F}.

In all applications, we employ the OU kernel $k_\rho(t,t^\prime) = \exp(-\frac{\|t-t^\prime \|}{\rho})$.  Concerning the offset prior $p(\bd{b})=\prod_{l=1}^L \mathcal{N}(b_l|\mu_{b_l},\sigma_{b_l}^2)$,
we employ a prior centred on the sample means $\bar{y}_l$ of the observed light curves, but with a highly inflated variance in order to render it only weakly informative, so that $\mu_{b_l} =  \bar{y}_l = \frac{1}{N_l}\sum_n^{N_l} y_l(t_{l,n})$ and $\sigma_{b_l}^2=100 \frac{1}{N_l}\sum_n^{N_l} (y_l(t_{l,n}) -\bar{y}_l)^2$. Without loss of generality, we fix $\tau_1 = 0.0$ and seek only the second delay $\tau=\tau_2$. In all applications, we consider candidate delays in the grid $\tau\in \{0.0, 0.2, \dots, 140\}$ of step size $0.2$ days.

\begin{figure}
    \centering
    \includegraphics[width=\columnwidth]{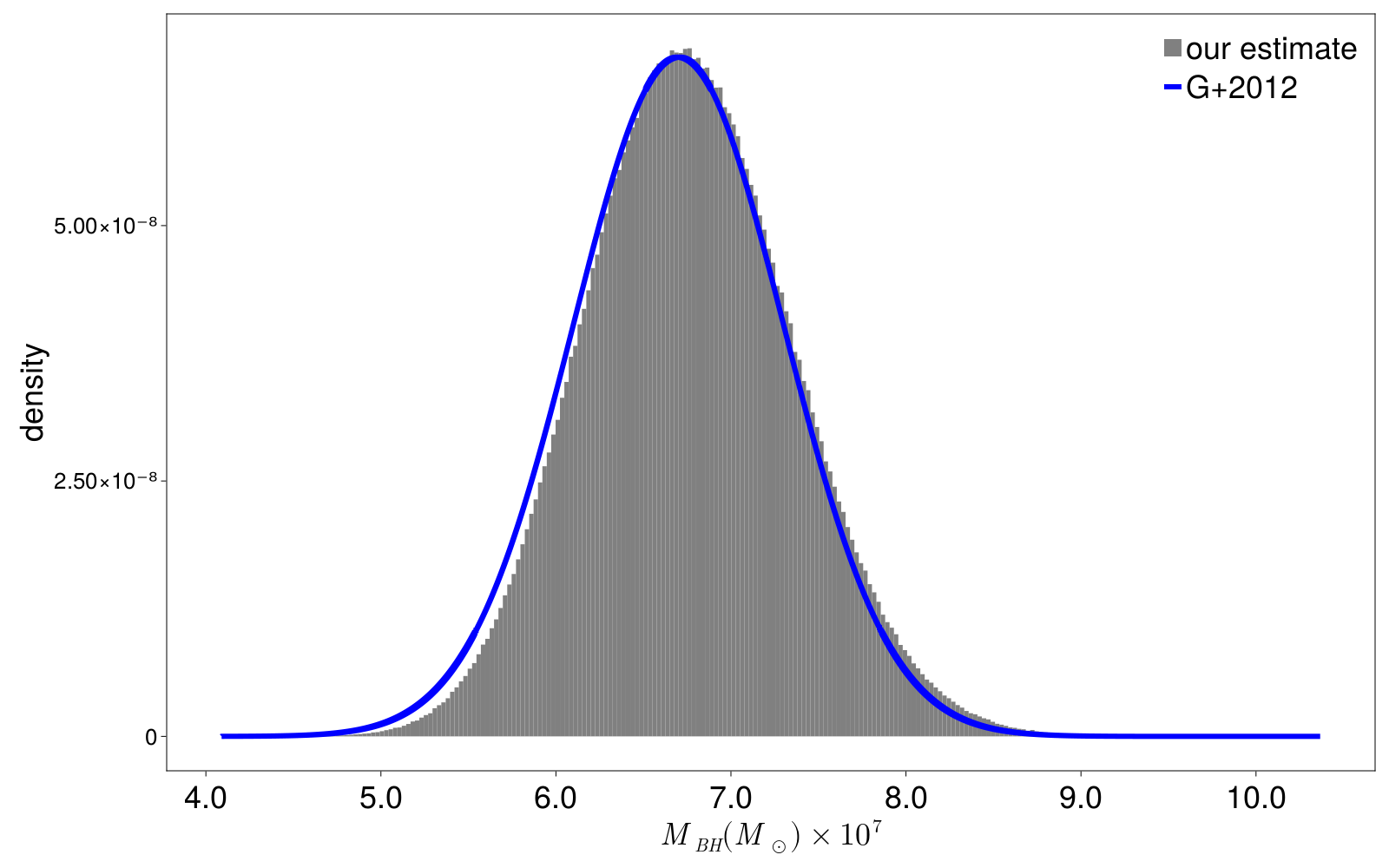}
    \caption{The grey histogram is our probability distribution estimate for the black hole mass of 3C120. The blue line corresponds to the $M_{\rm BH} = (6.7 \pm 0.6) \times 10^{7} M_{\odot}$ estimate reported in G+2012.}
    \label{fig:3C120Bhmass}
\end{figure}

Figure~\ref{fig:GPCCexamp} shows the results for source 3C120.
We present results for other sources  in appendix \ref{app:individualsources}.
The posterior distribution of the delay obtained with the prior $p(\tau)\propto 1$ (middle panel, Figure~\ref{fig:GPCCexamp}) broadly agrees with the ICCF (bottom panel, Figure~\ref{fig:GPCCexamp}).
Multiple peaks occur in the region between 10 and 40 days, consistent with the peak and centroid measured in the ICCF (values in bottom panel Figure~\ref{fig:GPCCexamp}, see also Table 8 in G+2012).
Higher probability peaks are observed at larger delays (> 100 days) may arise due to noise and the small number of overlapping points, as discussed in Section~\ref{sec3}.
A similar behaviour, that could be attributed to the same cause, is also observed in the ICCF for delays longer than $\sim80$ days.
We note that our range of the ICCF calculation is larger than that in G+2012 (their Figure 4) and we consider delays closer to the length of the light curves.
Based on the reported host subtracted luminosity of 3C120, $L_{5100\AA} = (9.12 \pm 1.15)\times 10^{43}{\mathrm{erg\ s^{-1}}}$ (see Table 9 in G+2012), we infer the posterior distribution of the delay using the prior $p(\tau)=\mathcal{U}(0,\tau_{max}(l,z))$ as defined in Section~\ref{sec:prior_on_delay} (top panel, Figure~\ref{fig:GPCCexamp}).
The introduction of the physically motivated prior (see \ref{sec:prior_on_delay}) suppresses all peaks at larger delays and accentuates the peaks at lower delays. The peak of highest probability occurs at $27.6$ days which is close to the ICCF peak and centroid estimates.

\begin{figure}
    \centering
    \includegraphics[width=\columnwidth]{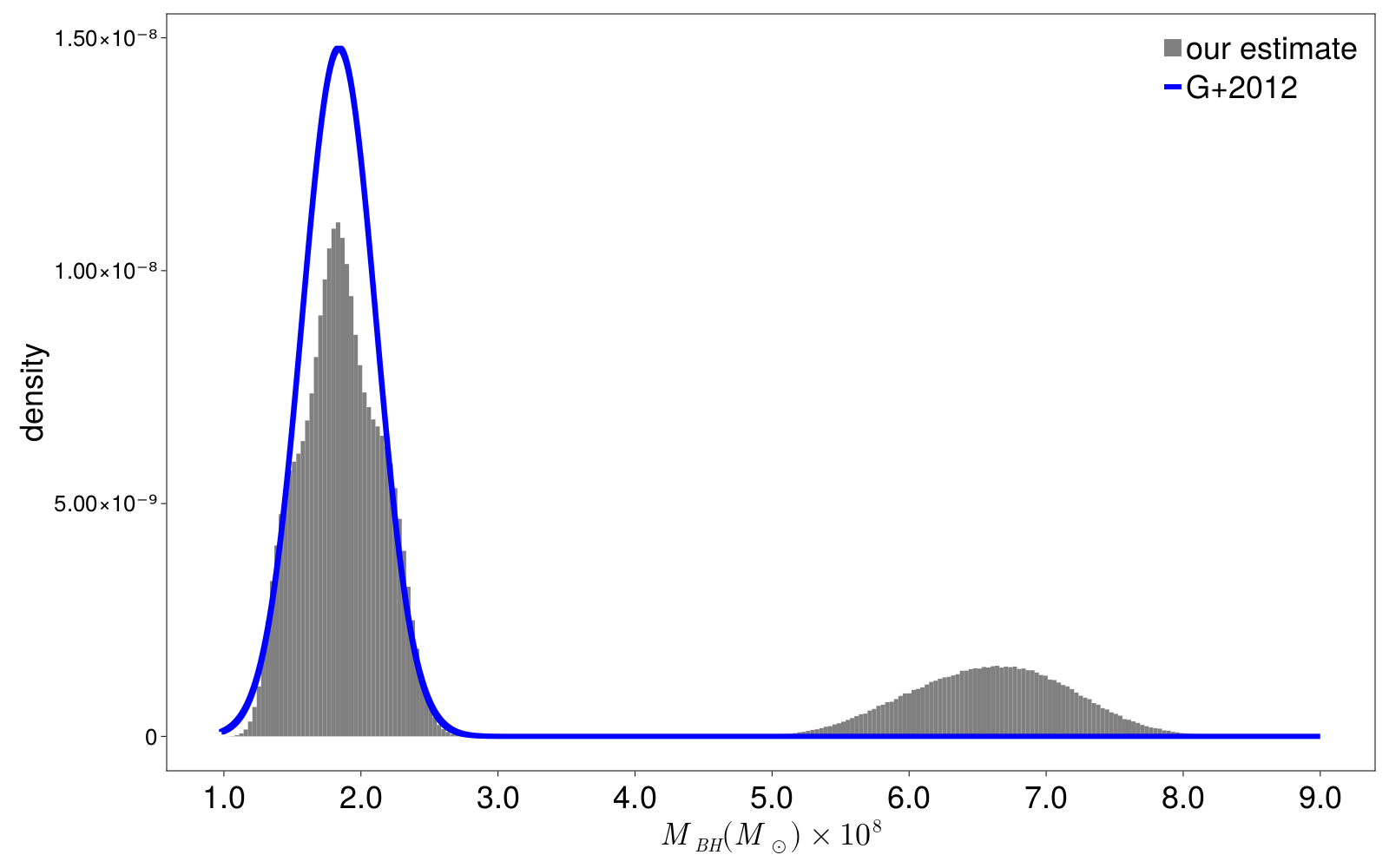}
    \caption{The grey histogram is our  probability distribution estimate for the black hole mass of Mrk1501. The blue line corresponds to the $M_{\rm BH} = (184 \pm 27) \times 10^{6} M_{\odot}$ estimate reported in G+2012.}
    \label{fig:Mrk1501Bhmass}
\end{figure}

\paragraph{Mass estimation for 3C120 and Mrk1501.} 
The posterior distribution captures all possible solutions for the delay.
This set of possible solutions can be incorporated into the calculation of other quantities. In the following, we use the recovered delay posterior (see top panel in Figure \ref{fig:GPCCexamp}), calculated with the prior presented in Section~\ref{sec:prior_on_delay},  to estimate the black hole mass distribution of 3C120 and Mrk1501.
According to the virial theorem, the black hole mass $M_{\rm BH}$ is given by
\begin{eqnarray}
{M_{\rm BH}} = f \frac{R \cdot \sigma_{V}^2}{G} \ ,
\label{eq:bhmvirial}
\end{eqnarray}
where $\sigma_{V}$ is the velocity dispersion of the emission lines, $R=c \cdot \tau$ is the size of the BLR, and the factor $f$ depends on the geometry and kinematics of the BLR (\citealt{2014A&A...568A..36P} and references therein).
As in G+2012, we use the velocity dispersion given by the linewidth of the RMS residual spectrum, $\sigma_{V} = 1514\pm 65 \, \rm{km/s}$ (their table 9), and assume a scaling factor $f=5.5$ (\citealt{2004ApJ...615..645O}).
We show the resulting probability distribution for the black hole mass in Figure~\ref{fig:3C120Bhmass}. We emphasize that  the distribution we calculate accounts for the uncertainty in the velocity dispersion $\sigma_{V}$.
G+2012 reported an estimate of $M_{\rm BH} = (6.7 \pm 0.6) \times 10^{7} M_{\odot}$ obtained by error propagation of the delay and velocity dispersion in Equation~\eqref{eq:bhmvirial}. As seen in Figure~\ref{fig:3C120Bhmass}, both estimates agree closely. We surmise, however, that this close agreement seems to be due to the variance of the velocity dispersion dominating the variance of the delay estimate. We provide further detail in Appendix \ref{app:mass3C120_fixdelaytomean}.

\begin{figure}
    \centering
    \includegraphics[width=0.9\columnwidth]{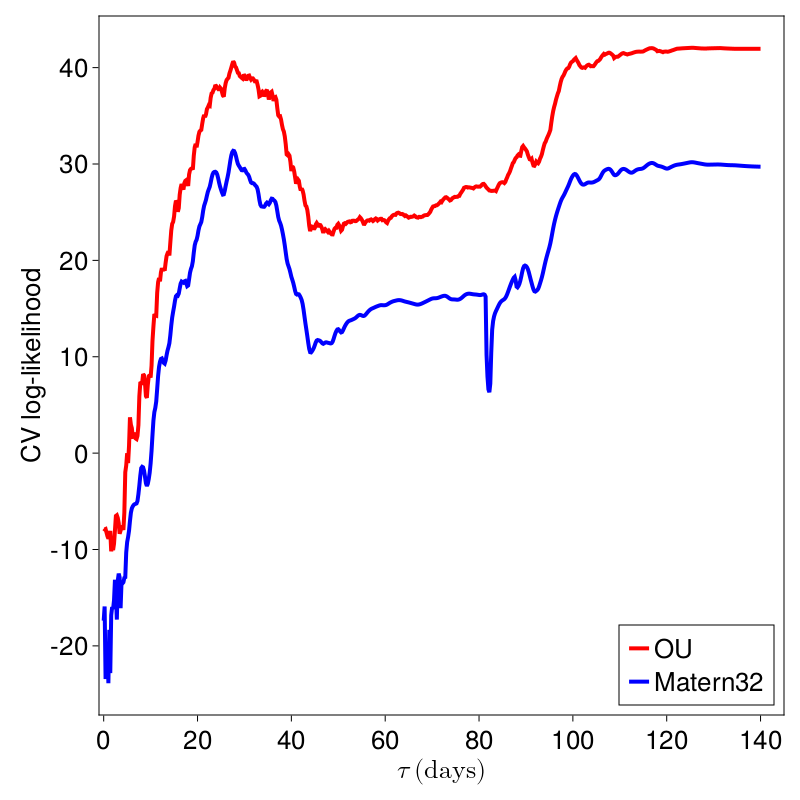}
    \caption{Cross-validation results for 3C120, OU vs Matern32.}
    \label{fig:3C120_CV}
\end{figure}

\begin{figure}
    \centering
    \includegraphics[width=0.9\columnwidth]{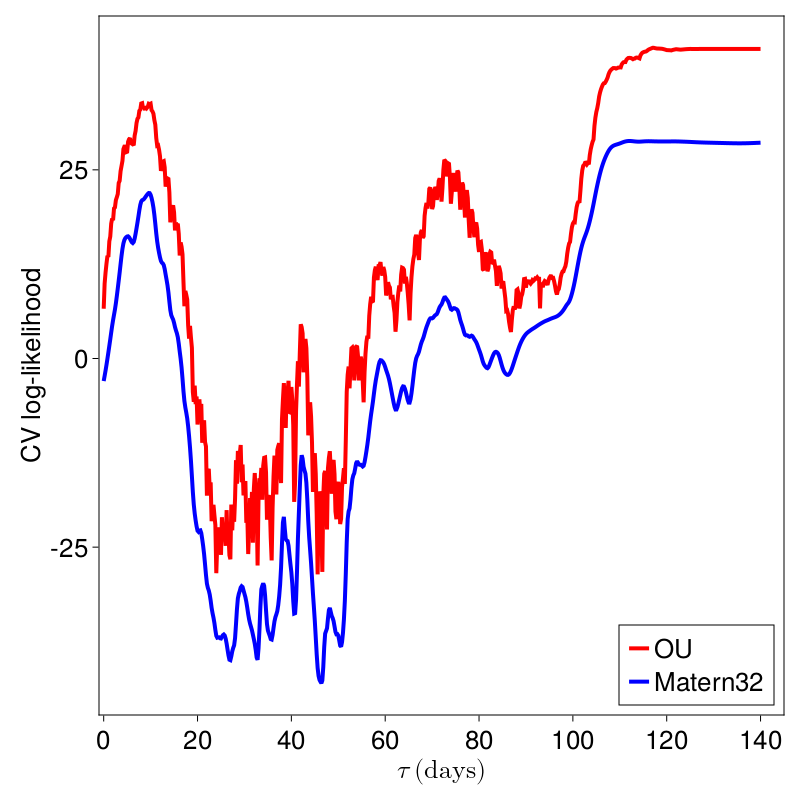}
    \caption{Same as Figure \ref{fig:3C120_CV} but for Mrk6.}
    \label{fig:Mrk6_CV}
\end{figure}

\begin{figure}
  \centering
  \includegraphics[width=1.0\columnwidth]{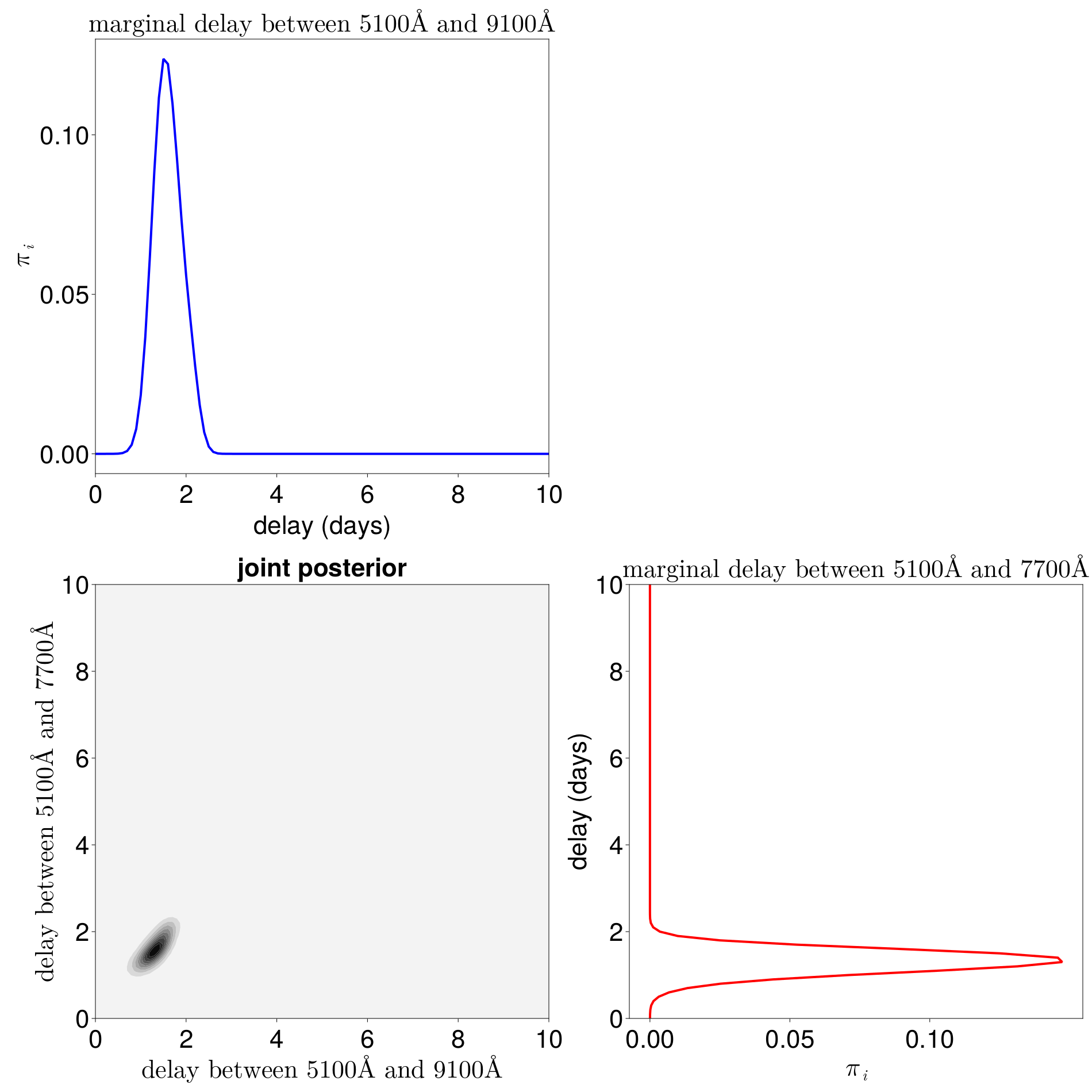}
  \includegraphics[width=1.0\columnwidth]{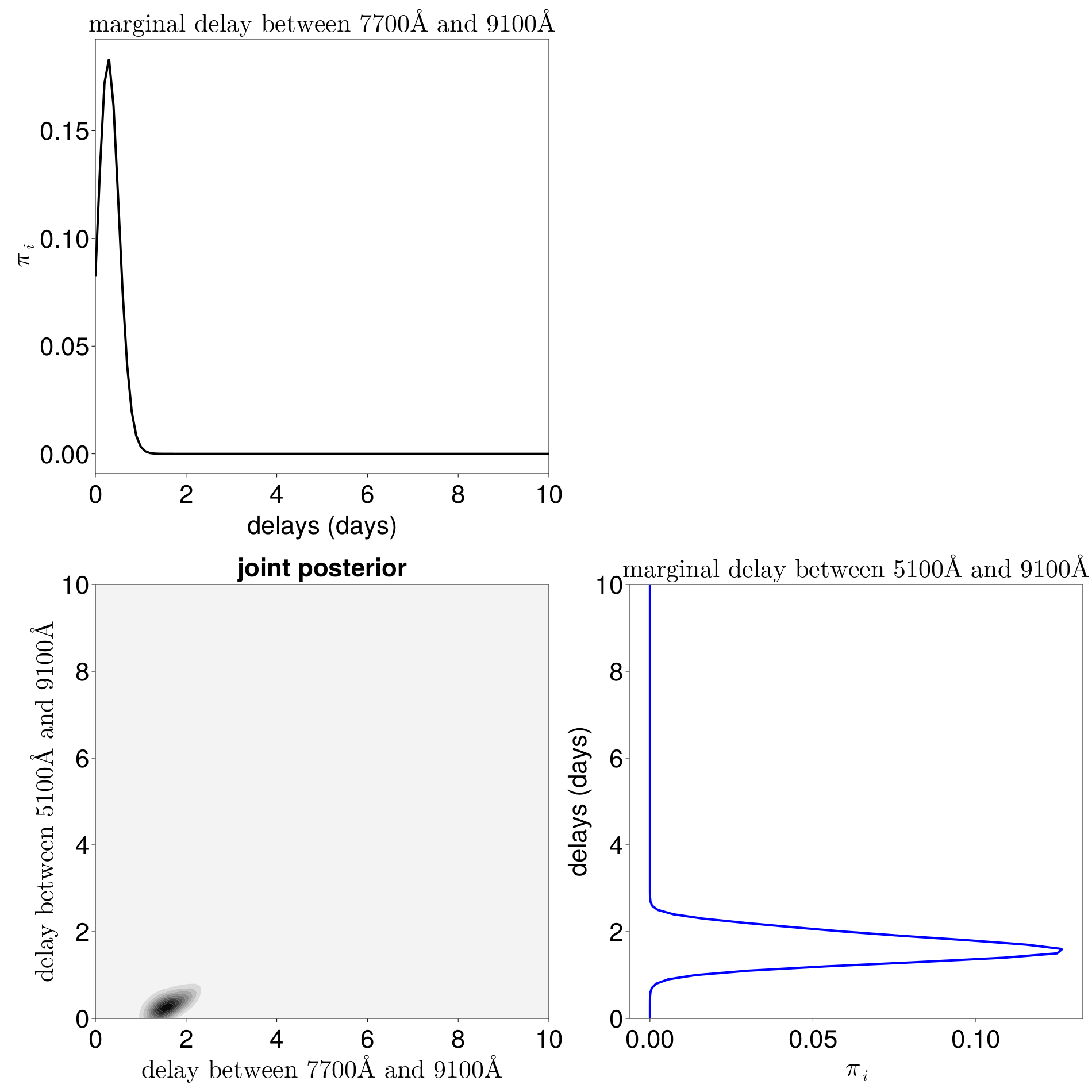}
  \caption{Joint delay posteriors for MGC+08-11-011.}
   \label{fig:Mgc0811_2D_posterior}
\end{figure}

\begin{figure}
  \centering
  \includegraphics[width=1\columnwidth]{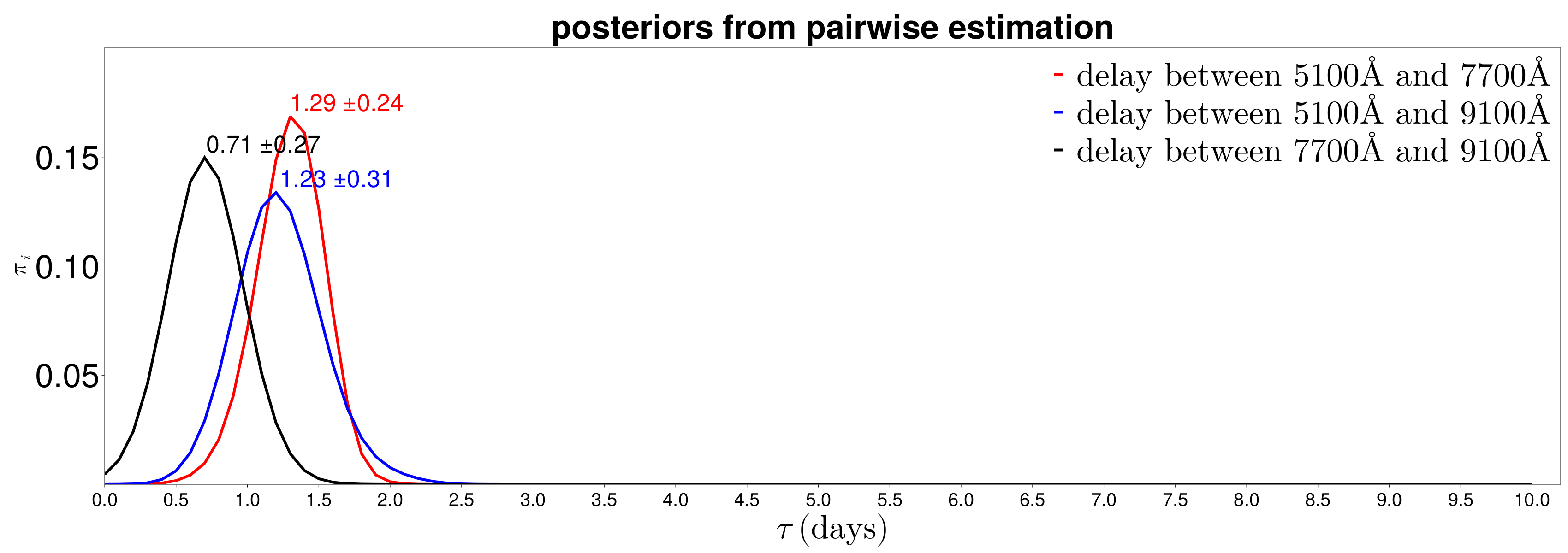}
  \includegraphics[width=1\columnwidth]{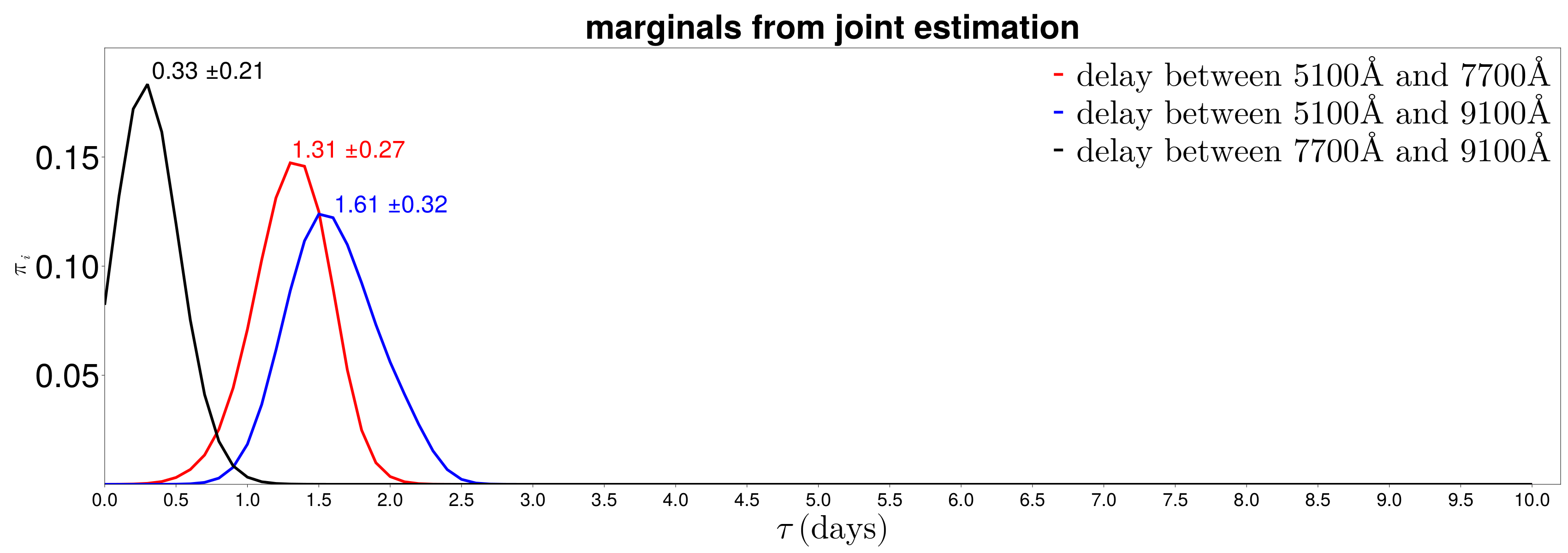}
  \caption{Pairwise (top) and joint (bottom) estimation of delays. Each density is given its mean and standard deviation. We note that the two estimation methods give different results when the delays are estimated for three light curves in MGC +08-11-011.}
  \label{fig:pairwise_vs_joint}
\end{figure}

In Figure~\ref{fig:Mrk1501Bhmass} we show the probability distribution for the black hole mass of Mrk1501. Again we use \eqref{eq:bhmvirial}, with values $f=5.5$ and $\sigma_{V} = 3321\pm 107 \, \rm{km/s}$ and the corresponding delay posterior (see top panel in Figure \ref{fig:GPCCexampMrk1501}), to calculate the mass distribution. The resulting distribution indicates that potential candidate values for the black hole mass concentrate around two distinct modes. G+2012 reported an estimate of $M_{\rm BH} = (184 \pm 27) \times 10^{6} M_{\odot}$, which is in good agreement with one of the modes of our estimate.

As mentioned above, the variance in velocity dispersion dominates the variance in delay when calculating the black hole mass distribution\footnote{We note that the geometry effects contained in the scaling factor $f$ are the main source of uncertainty in the estimates of the black hole mass from RM.}. For high redshift quasars, for example, the uncertainty in the time delay increases considerably, mainly due to significant seasonal gaps and the decrease in the amplitude of the variability. This poses an additional challenge when using delay-luminosity scaling relations (e.g. based on the CIV line) to estimate black hole masses from a single- epoch spectrum (e.g., \citealt{2021ApJ...915..129K}; \citealt{2019ApJ...887...38G}).
One could use very high-resolution spectroscopy ($R\sim 20000$) to improve the accuracy of the velocity dispersion to $\sim 1$\% (e.g. \citealt{2020MNRAS.493.3656C}). Obviously, under such conditions, the uncertainty in the delay in equation \ref{eq:bhmvirial} becomes relatively more influential and thus an improved description of the delay posterior distribution, as delivered by the GPCC, is important.

\paragraph{Model selection.} The GPCC provides out-of-sample predictions (see Equation \eqref{eq:predictive_likelihood}) and can thus be subjected to model selection via cross-validation (CV). This allows us not only to compare GPCC to other models in terms of predictive performance, but also to validate certain choices when using GPCC, such as the kernel choice. We briefly demonstrate how CV allows us to decide which one of two candidate kernels\footnote{Alternative sets of kernels can be included in the model selection.}, OU or Matern32, provides a better fit for the datasets 3C120 and Mrk6. We consider delays in $\{0.0, 0.2, \dots, 140\}$. We perform $10$-cross fold validation for each candidate delay. Figures \ref{fig:3C120_CV} and \ref{fig:Mrk6_CV} display the cross-validated log-likelihood for the two kernels. Evidently, the kernels perform similarly, but the OU kernel shows consistently better performance and is thus the better choice of the two. We note that this kind of model selection is not possible in the case of the ICCF --- for instance, we may want to choose between two ways of interpolating the observed fluxes --- as it does not provide out-of-sample predictions.

\paragraph{Three lightcurves, MGC+08-11-011.} As an example of the analysis of more than two light curves, we use the data published in \cite{2018ApJ...854..107F} for the source MCG+08-11-011 as part of a RM campaign of the accretion disc. We use the light curves at bands $5100$\,\AA, and the Sloan $i$ and $z$ with central wavelengths at $7700$\,\AA, and $9100$\,\AA, respectively. We consider delays in $\{0,0.1,\dots, 10\}$. Here, the maximum delay of $10$ days is well above what standard accretion disc theory predicts (e.g., \citealt{2019MNRAS.490.3936P, 2022arXiv221209161P}). The joint posterior is shown in Figure \ref{fig:Mgc0811_2D_posterior}. 
The orientation of the joint posterior distribution reveals a positive correlation between the delays.

As pointed out in Section \ref{sec:morethantwolightcurves}, this is a case where GPCC has an advantage over ICCF as the latter cannot produce a joint estimate of the two sought delays, but must instead consider pairs of light curves at a time. In fact, this pairwise estimation may lead to an inconsistency. To demonstrate this inconsistency, we use the GPCC to estimate the delays between the three pairs of lightcurves $5100$\AA\ and $7700$\AA, $5100$\AA\ and $9100$\AA, $7700$\AA\ and $9100$\AA , as opposed to the joint estimation that we show above. We display the results of this pairwise estimation in Figure \ref{fig:pairwise_vs_joint} (top). The results tell us that the mean delay between $5100$\AA\ and $7700$\AA\ is $1.29$ days (red), and that the mean delay  between 
$5100$\AA\ and $9100$\AA\ is $1.23$ days (blue). Given these two pieces of information, we would expect the mean delay between $7700$\AA\ and $9100$\AA\ to be approximately\footnote{We note that comparing the means of the distributions of delays is only an approximate way of checking whether the estimated delays are consistent with one another.} $1.23-1.29=-0.06$ days, but the pairwise comparison yields a mean of $0.71$ days (black). Hence, the pairwise estimation leads to an inconsistency as the distribution of the pairwise delays suggests that the mean delay for $7700$\AA\ and $9100$\AA\ should be closer to $0.06$ days rather than the independently estimated $0.71$ days.
For comparison, we display in Figure \ref{fig:pairwise_vs_joint} (bottom) the marginals calculated from the joint posterior in Figure \ref{fig:Mgc0811_2D_posterior}. The figure shows that the mean delay between $5100$\AA\ and $7700$\AA\ is $1.31$ days (red), and that the mean delay between 
$5100$\AA\ and $9100$\AA\ is $1.61$ days (blue) which is consistent with the mean delay of $0.33$ (black) between $7700$\AA\ and $9100$\AA\ .

\section{Summary and conclusions}\label{sec5}

We have presented a probabilistic reformulation of the ICCF method to estimate time delays in RM of AGN.
The main features and advantages of our method can be summarised as follows:

\begin{itemize}

      \item It accounts for observational noise and provides a posterior probability density of the delay. The ICCF provides only a point estimate for the delay. A distribution for the delay can be obtained using the bootstrapping method as implemented in FR/RSS, but it has the disadvantage of being susceptible to the removal of individual data points. The posterior distribution obtained in the Bayesian framework is a powerful alternative to bootstrapping.

      \item When using the ICCF, one has to manually choose an object-dependent threshold to determine the peak and centroid of the cross-correlation curve. The proposed model avoids such manual choices by relying on a probabilistic formulation that defines a cost function, the marginal log-likelihood. This guides the optimisation of the parameters and is used in the derivation of the posterior density for the delay.

      \item The proposed model can jointly estimate the delay between more than two light curves, a task that cannot be accomplished with the ICCF. The ICCF follows a
      pairwise estimation approach that can lead to delays that are inconsistent with each other.

      \item  Finally, we have demonstrated that the capability of the proposed model for out-of-sample predictions allows us to use cross-validation for choosing the kernel. Similarly, we could have used cross-validation to call into question other design choices of the model such as the noise model. The same capability would also allow us to pit our proposed model against other competing models in a cross-validation framework in order to decide which one fits the observed data at hand the best. The ICCF does not possess this capability.
      
\end{itemize}

\begin{acknowledgements}

The authors gratefully acknowledge the generous and invaluable support of the Klaus Tschira Foundation. 
FPN warmly thank Bozena Czerny for discussions.
This project has received funding from the European Research Council (ERC) under the European Union's Horizon 2020 research and innovation programme (grant agreement No 951549).
This research has made use of the NASA/IPAC Extragalactic Database (NED) which is operated by the Jet Propulsion Laboratory, California Institute of Technology, under contract with the National Aeronautics and Space Administration.
This research has made use of the SIMBAD database, operated at CDS, Strasbourg, France.
We thank the anonymous referee for constructive comments and careful review of the manuscript.
      
\end{acknowledgements}

\begin{appendix}

\section{Matrix Q}
\label{app:matrix_q}

Matrix $\bd{Q}$ has dimensions $N\times L$. Its $l$-th column reads:
\begin{align*}
l=1, \ &(\underbrace{1,\dots,1}_{N_1}, 0, \dots, 0) \ ,\\
&\vdots \\
l=\ell, \ &(0,\dots,0,\underbrace{1,\dots,1}_{N_\ell}, 0 \dots, 0) \ , \\
&\vdots \\
l=L, \ &(0,\dots,0,\underbrace{1,\dots,1}_{N_L}) \ .
\end{align*}
We construct the columns of the $N^*\times L$ matrix $\bd{Q}^*$, used when calculating the predictive likelihood in Equation~\eqref{eq:predictive_likelihood}, in the exact same fashion.

\section{Predictive likelihood}
\label{app:predictive_likelihood}

The joint distribution of observed fluxes and new fluxes reads:
\begin{align}
p(\bd{y},\bd{y}^*|\bd{\tau},
\bd{\alpha}, \rho) &= \notag\\ 
&\mathcal{N}\bigg(
\begin{pmatrix} 
\bd{y} \\ \bd{y}^* \end{pmatrix} 
| 
\begin{pmatrix} 
\bd{Q}\bd{b} \\ \bd{Q}^*\bd{b} \end{pmatrix},
\begin{pmatrix} 
\bd{C}(\bd{t},\bd{t}) && \bd{C}_*(\bd{t},\bd{t}^*) \\ \bd{C}_*(\bd{t},\bd{t}^*)^T && \bd{C}(\bd{t}^*,\bd{t}^*)  \end{pmatrix}
\bigg)
\label{eq:joint_train_test}
\end{align}
By conditioning on \bd{y}, we obtain Equation \eqref{eq:predictive_likelihood} from Equation \eqref{eq:joint_train_test}.

\section{Simulations}
\label{app:simulations}

\begin{figure}
    \centering
    \includegraphics[width=\columnwidth]{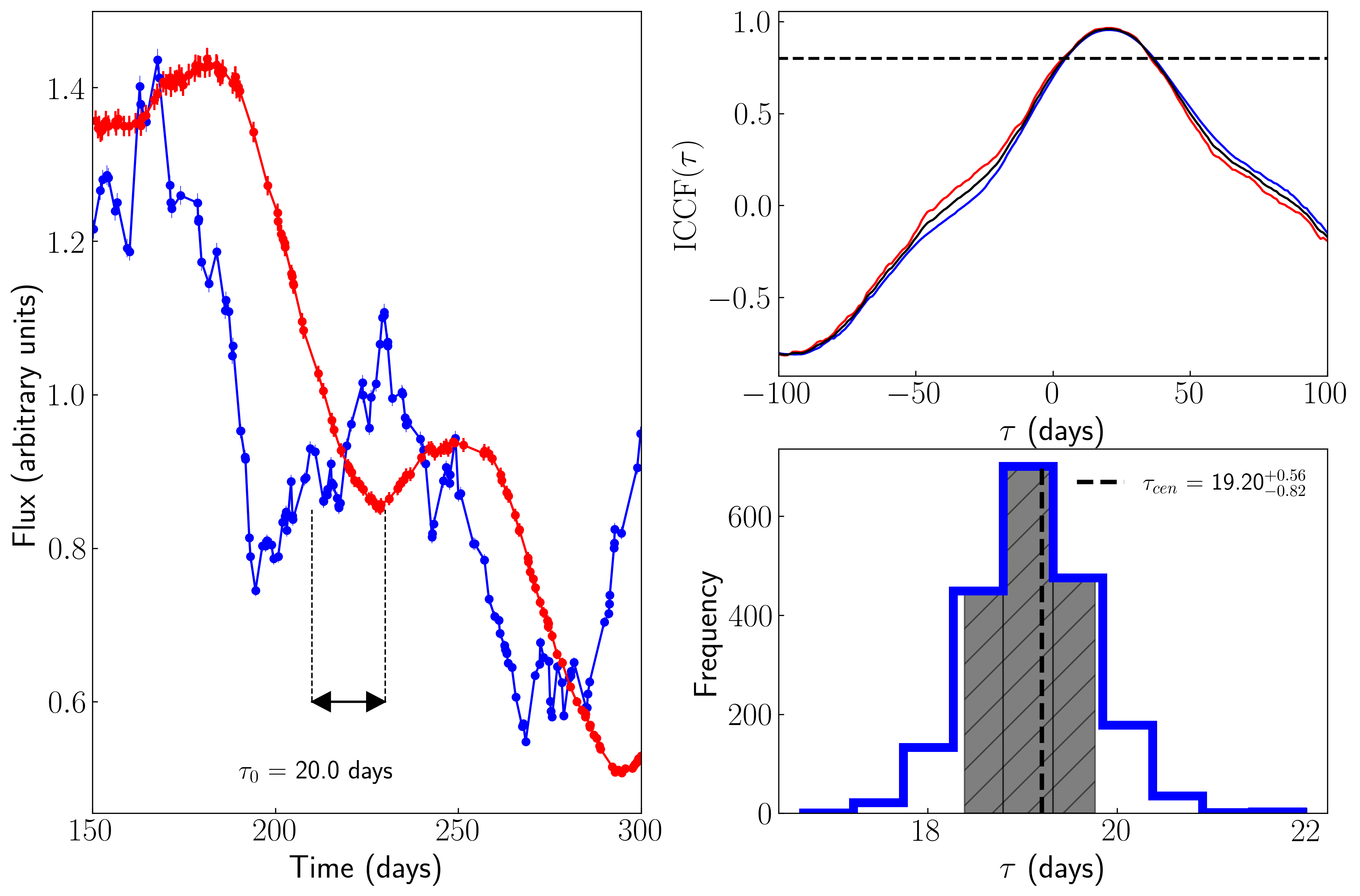}
    \caption{ICCF application to simulated data.
    \emph{Left panel}: AD continuum light curve (blue) modeled as a random walk process with a power spectral density $P(\nu)\propto \nu^{-2}$ (\citealt{2009ApJ...698..895K}; \citealt{2017ApJ...834..111C}).
    The delayed BLR emission line light curve (red) is obtained by convolving the AD light curve with a rectangular transfer function over the time domain, $t \in \bigr[\tau_{0} - \Delta \tau/2, \tau_{0} + \Delta \tau/2 \bigr]$ where $\Delta \tau/\tau_{0} \leq 2$ (\citealt{2013A&A...552A...1P}).
    For this particular example, a centroid $\tau_{0} = 20$ days is assumed.
    Both light curves were randomly sampled, with an average sampling of 1 day and noise level of 1\% (S/N = 100).
    \emph{Right Upper Panel}:
    ICCF results when the line is interpolated and the continuum is shifted (blue) and when the continuum is interpolated and the line is shifted (red). The average ICCF is shown in black.
    The dotted black line marks the threshold of $0.8max$(ICCF) used to calculate the centroid.
    \emph{Right bottom panel}: The histogram shows the distribution of the centroid delay obtained by the FR/RSS method (see text).
    The black area marks the 68\% confidence range used to calculate the errors of the centroid.}
    \label{fig:iccfexamp}
\end{figure}

\begin{figure*}
    \centering
    \includegraphics[width=18cm,clip=true]{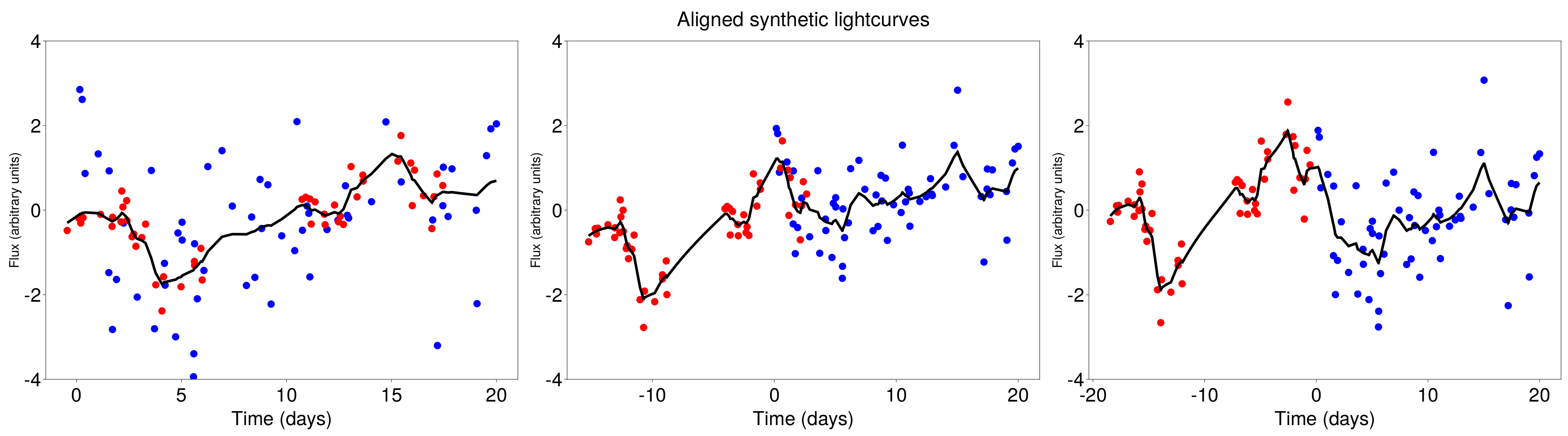}
    \caption{Synthetic light curves for $\sigma=1.5$ aligned with delays $2$ (left), $16.8$ (middle) and $20$ (right). Vertical axis is in arbitrary units of simulated flux. The black line is the recovered latent signal $f(t)$ that underlies all observations (see \ref{sec:model_formulation}). All three delays lead to plausible alignments.}
    \label{fig:GPCCexampsimul}
\end{figure*}

Figure \ref{fig:iccfexamp} shows an example of the ICCF application to simulated data.
Figure \ref{fig:GPCCexampsimul} shows how the synthetic curves, considered in section \ref{sec3}, align for candidate posterior delays for the case of $\sigma=1.5$.

\section{Individual sources}
\label{app:individualsources}

\begin{figure}
    \centering
    \includegraphics[width=\columnwidth]{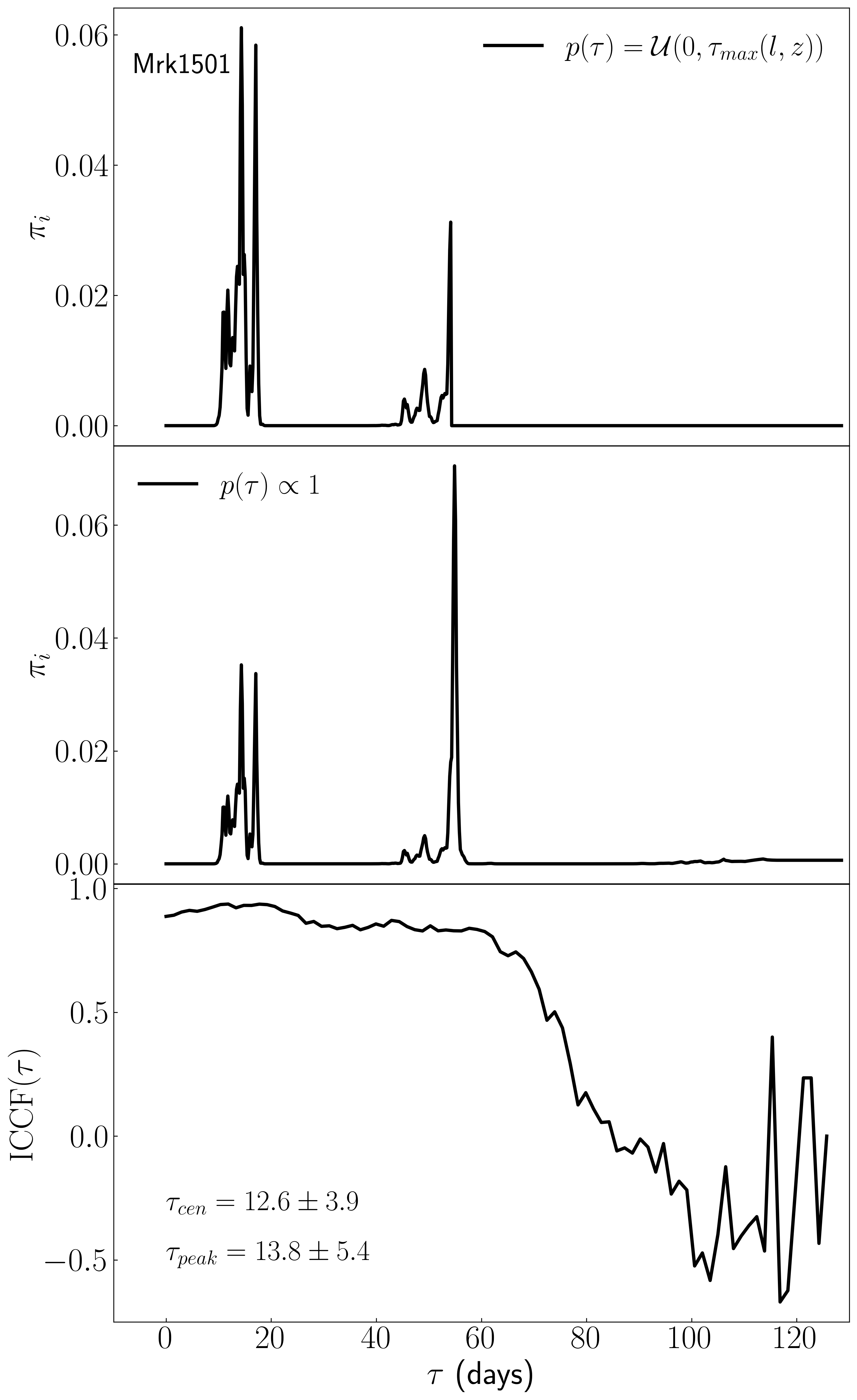}
    \caption{same as Figure \ref{fig:GPCCexamp} but for Mrk1501.}
    \label{fig:GPCCexampMrk1501}
\end{figure}

\begin{figure}
    \centering
    \includegraphics[width=\columnwidth]{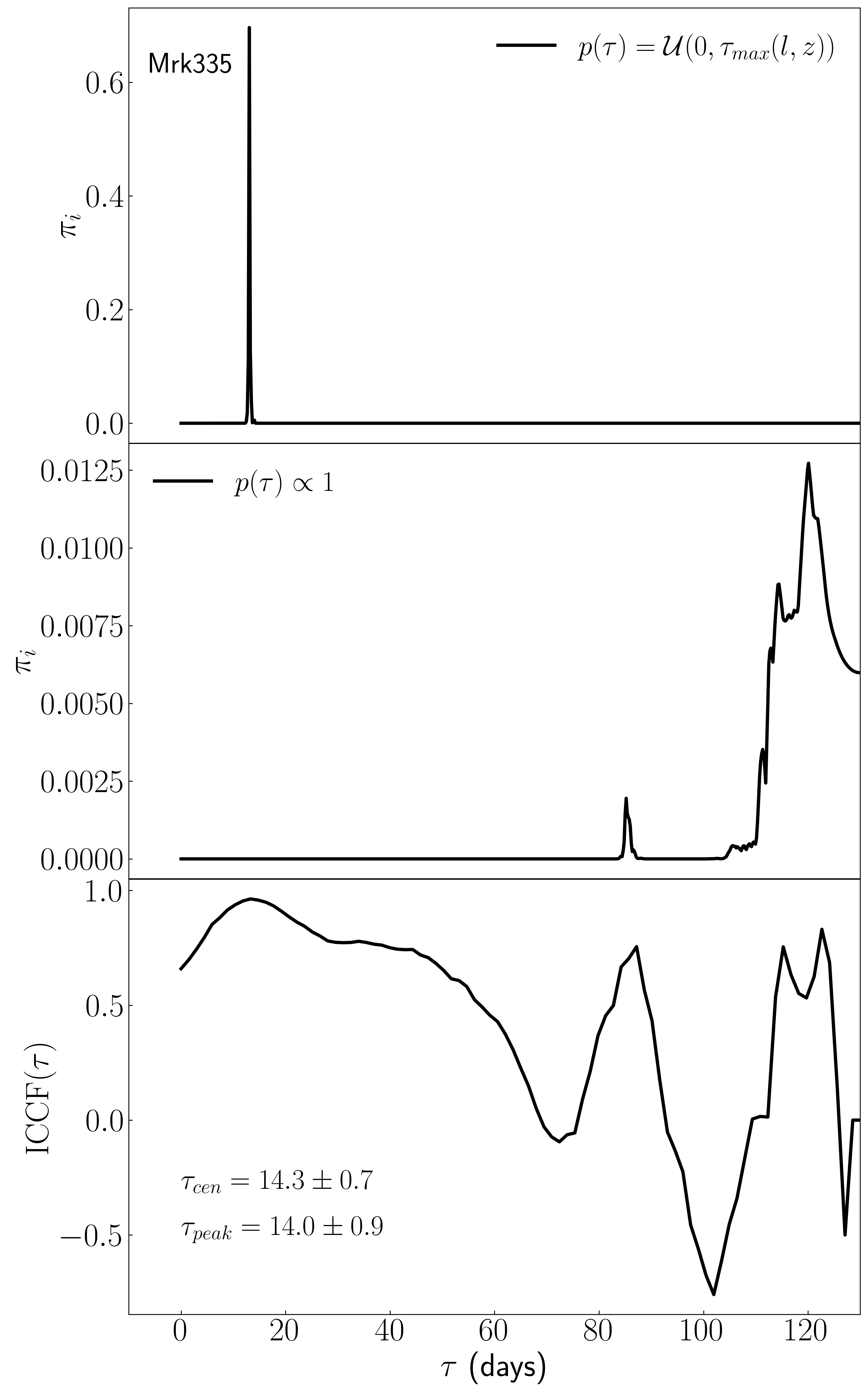}
    \caption{same as Figure \ref{fig:GPCCexamp} but for Mrk335.}
    \label{fig:GPCCexampMrk335}
\end{figure}

\begin{figure}
    \centering
    \includegraphics[width=\columnwidth]{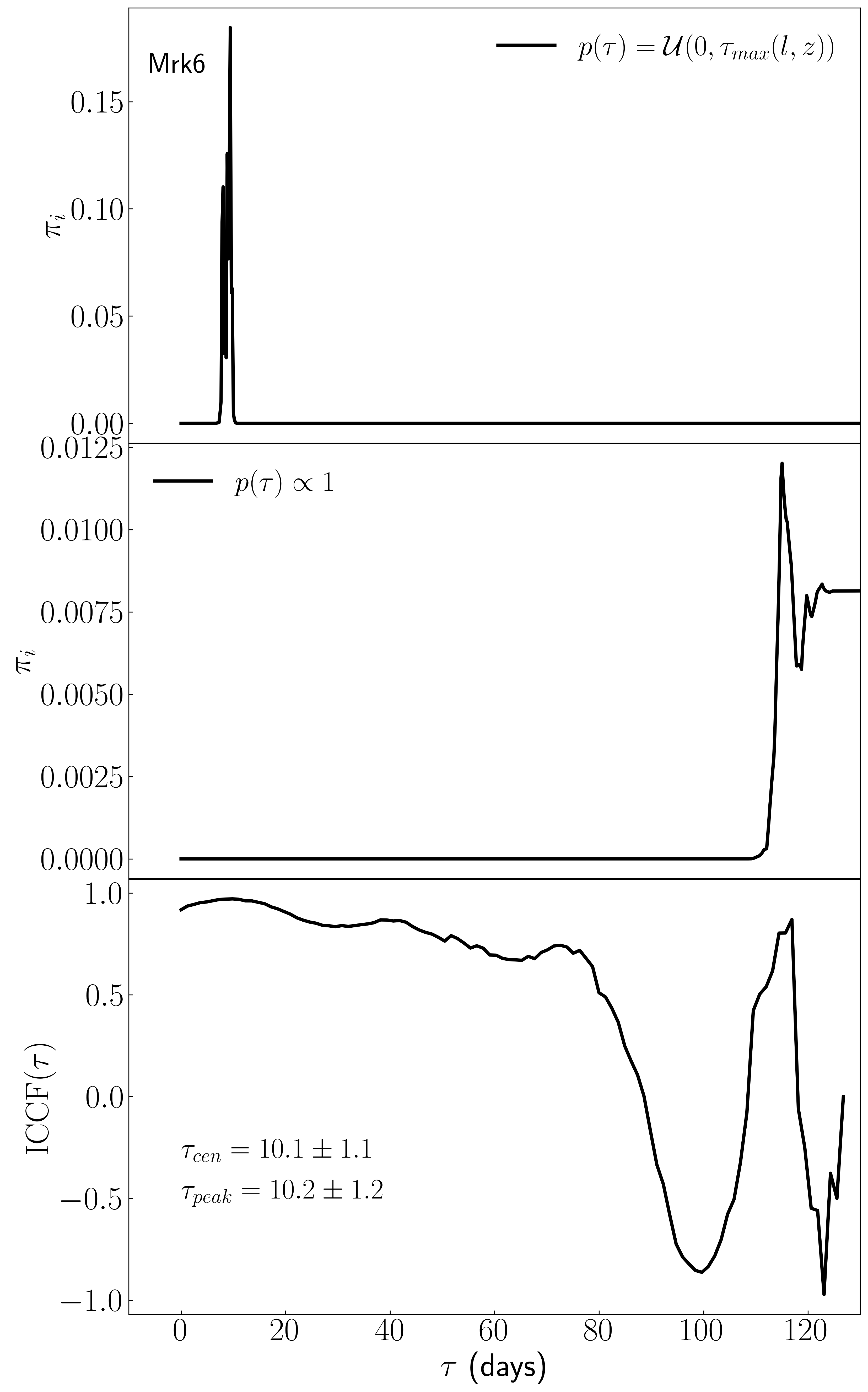}
    \caption{same as Figure \ref{fig:GPCCexamp} but for Mrk6.}
    \label{fig:GPCCexampMrk6}
\end{figure}

\begin{figure}
    \centering
    \includegraphics[width=\columnwidth]{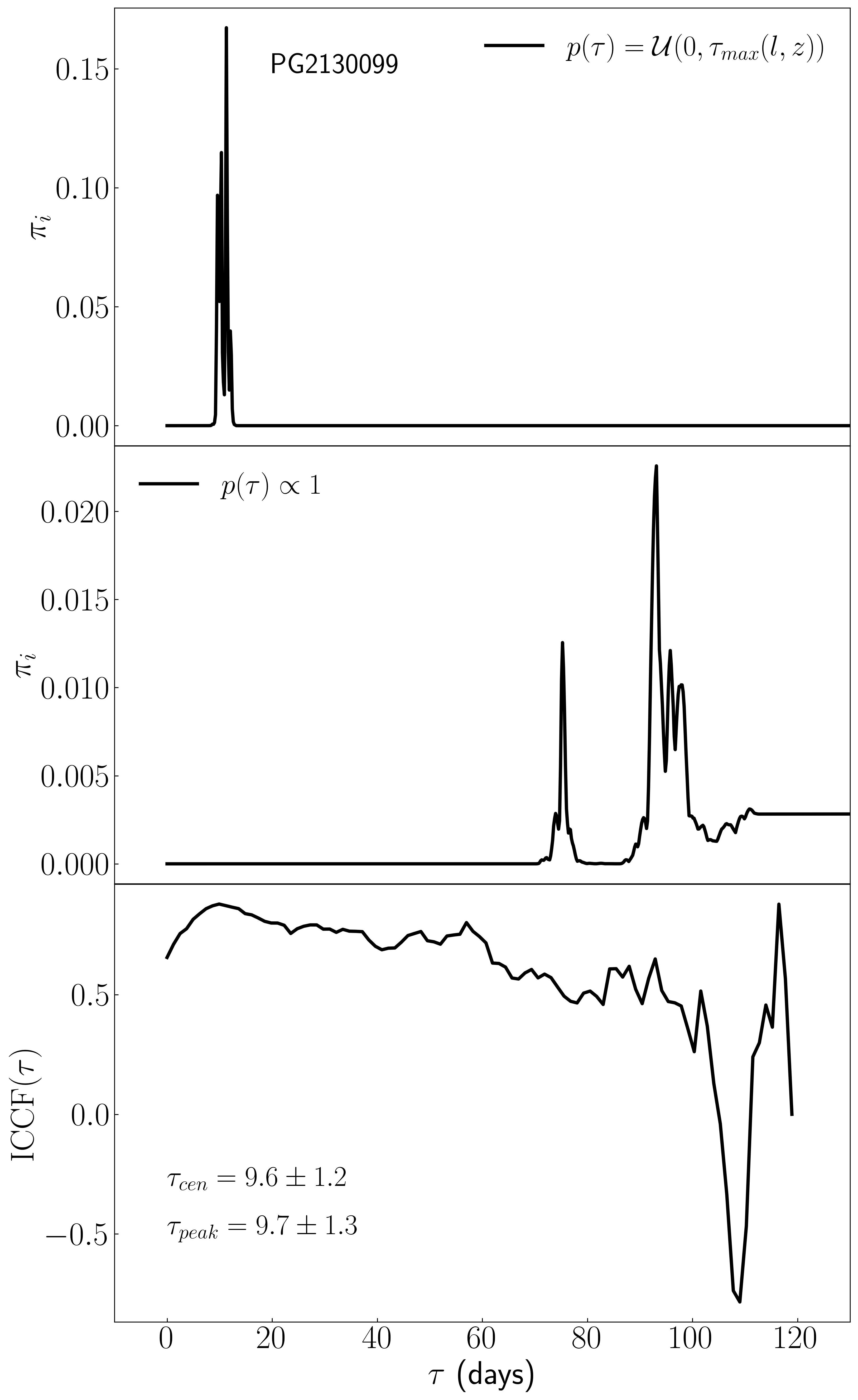}
    \caption{same as Figure \ref{fig:GPCCexamp} but for PG2130+099.}
    \label{fig:GPCCexampPG2130099}
\end{figure}

\section{Mass distribution of 3C120 for fixed delay}
\label{app:mass3C120_fixdelaytomean}

\begin{figure}
    \centering
    \includegraphics[width=\columnwidth]{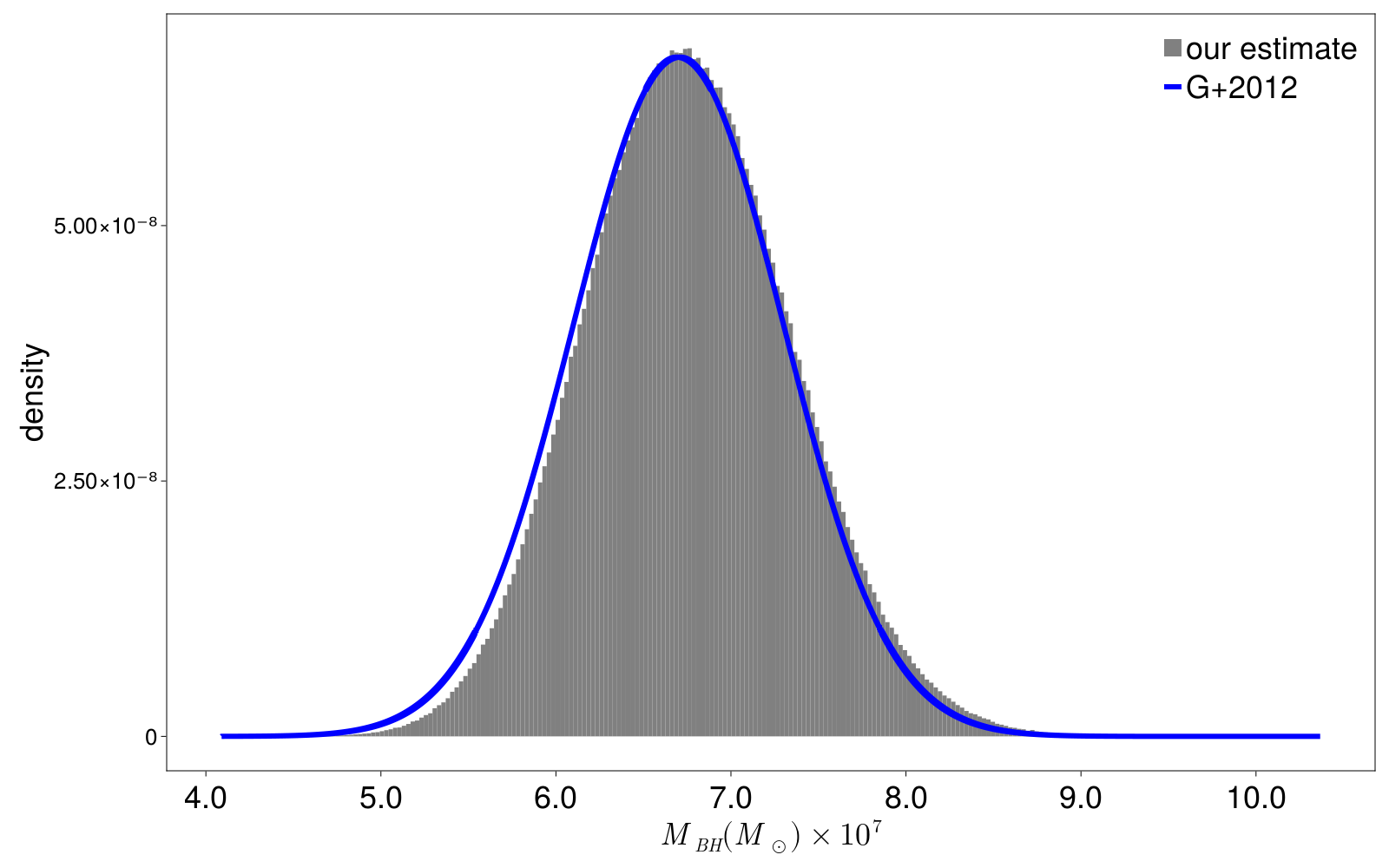}
    \caption{The grey histogram is our  probability density estimate for the black hole mass of 3C120 when fixing the delay $\tau$ to its posterior mean value of $27.4$ days. The observed variance of the grey histogram is entirely due to the variance of the velocity dispersion since delay $\tau$ is fixed to its mean posterior value. The blue line corresponds to the Gaussian probability density $M_{\rm BH} = (6.7 \pm 0.6) \times 10^{7} M_{\odot}$ reported in G+2012.}
    \label{fig:3C120Bhmass_fixdelaytomean}
\end{figure}
 We display an additional figure that shows the mass distribution of 3C120 calculated using Equation~\eqref{eq:bhmvirial} but with the delay $\tau$ fixed to its mean posterior value of $27.4$ and allowing only the velocity dispersion to vary according its Gaussian distribution implied by $\sigma_{V} = 1514\pm 65 \, km/s$. We note that Figure \ref{fig:3C120Bhmass_fixdelaytomean} is almost identical to Figure \ref{fig:3C120Bhmass}. The standard deviation of our distribution (grey histogram) for fixed delay in Figure \ref{fig:3C120Bhmass_fixdelaytomean} is $0.580\times10^7$, while the standard deviation of our distribution in  Figure \ref{fig:3C120Bhmass} (again grey histogram) has an only slightly larger value of $0.596\times10^7$. This suggests that the variance of the velocity dispersion dominates the variance of the delay. We surmise that this is the reason that makes our estimate and the G+2012 estimate appear in close agreement even though the corresponding black hole mass estimates might agree less. 
\end{appendix}

\end{document}